\documentclass[3p,12pt]{elsarticle}
\usepackage[fleqn]{amsmath}
\usepackage{cases}
\usepackage{booktabs}
\usepackage{multirow}
\usepackage{xcolor}
\usepackage[normalem]{ulem}
\usepackage{tabularx}
\usepackage{siunitx}

\usepackage{comment}
\usepackage[ruled, linesnumbered]{algorithm2e}
\usepackage{algpseudocode}
\usepackage{grffile}
\usepackage{rotating}
\usepackage{lineno}
\usepackage{hyperref}
\usepackage{graphicx,enumerate}
\usepackage{amssymb}
\allowdisplaybreaks[4]
\usepackage{bm}
\usepackage[caption=false]{subfig}
\usepackage{caption,color}
\usepackage{subeqnarray}
\usepackage{multirow}
\usepackage{lscape}
\usepackage{rotating}
\usepackage{graphics}
\usepackage{epstopdf}
\usepackage{threeparttable,booktabs,tabularx}
\journal{}
\usepackage{floatrow}
\usepackage{floatrow}\newfloatcommand{capbtabbox}{table}[][\FBwidth]
\floatname{algorithm}{Algorithm}
\biboptions{numbers,sort&compress} 

\begin{document}

\begin{frontmatter}
\title{Multiscale simulation of neutral particle flows in the  plasma edge }

\author{Yifan Wen}
\author{Yanbing Zhang}
\author{Lei Wu\corref{mycorrespondingauthor}}
\cortext[mycorrespondingauthor]{Corresponding author}
\ead{wul@sustech.edu.cn}

\address{Department of Mechanics and Aerospace Engineering, Southern University of Science and Technology, Shenzhen 518055, China}

\begin{abstract}
The plasma edge flow, situated at the intricate boundary between plasma and neutral particles,  plays a pivotal role in the design of nuclear fusion devices such as divertors and pumps. Traditional numerical simulation methods, such as the direct simulation Monte Carlo approach and the discrete velocity method, are hindered by extensive computation times when dealing with near-continuum flow conditions.
This paper presents a general synthetic iterative scheme to deterministically simulate the plasma edge flows. By alternately solving the kinetic equations and macroscopic synthetic equations, our method substantially decreases the number of iterations, while maintains asymptotic-preserving properties even when the spatial cell size is much larger than the mean free path. 
Consequently,  our approach achieves rapid convergence and high accuracy in plasma edge flow simulations, particularly in near-continuum flow regimes. This advancement provides a robust and efficient computational tool, essential for the advancement of next-generation nuclear fusion reactors.  
\end{abstract}


\end{frontmatter}

\section{Introduction}

\begin{figure}[t]
    \centering
    \includegraphics[width=0.37\linewidth]{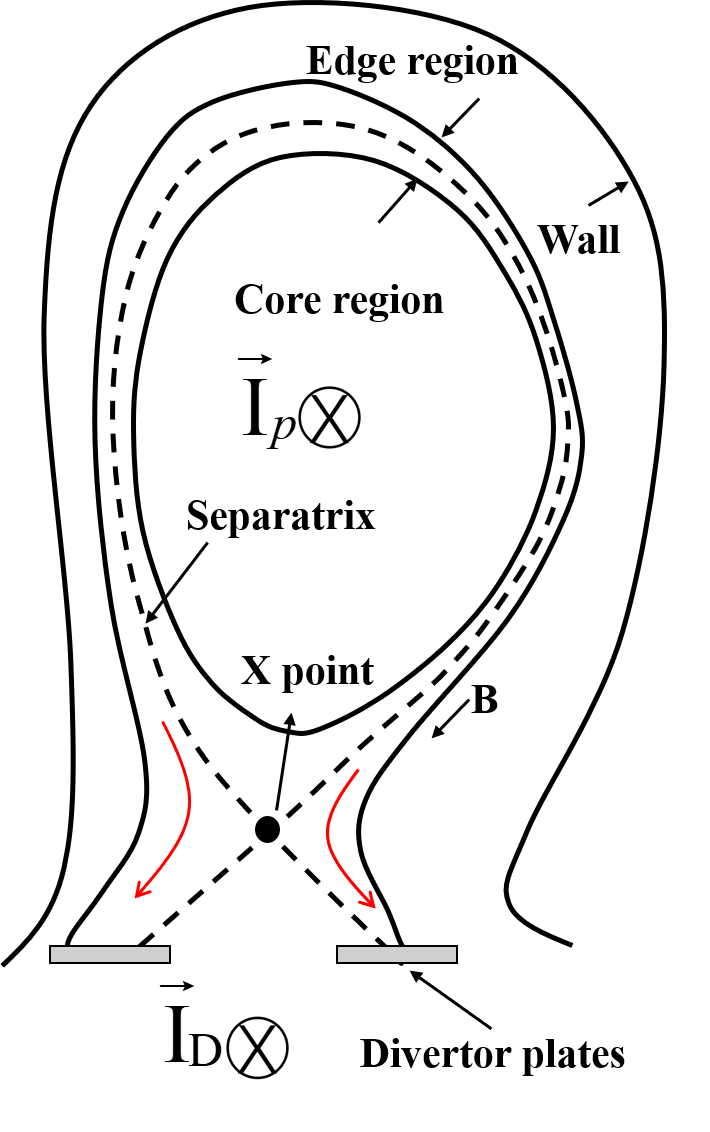}
    \caption{A schematic of the edge region in the poloidal plane of a divertor tokamak.}
    \label{fig:Edge_sketch}
\end{figure}

The edge plasma generally refers to the outer layer of plasma in fusion devices (such as tokamaks), consisting of a mixture of plasma and neutral particles \cite{HorstenJCP2020}. As shown in Fig.~\ref{fig:Edge_sketch}, the innermost layer is the inner wall surface of the device, with the majority of the poloidal magnetic field generated by the plasma current $I_p$, approximating a circular shape. An external coil with a current $I_D$ is aligned in the same direction as $I_p$ to create the so-called ``poloidal divertor'' structure. The magnetic fields produced by $I_D$ and $I_p$ cancel each other at the X-points. The magnetic surface passing through an X-point is called the separatrix. A solid plate intercepts the magnetic surfaces around the current $I_D$, transporting the high-temperature plasma from the core region to this solid plate via the magnetic field, and preventing direct contact with the inner wall. The region between the outer magnetic flux surface of the core and the separatrix is referred to as the edge plasma region \cite{Rognlien_2005,Blommaert_CPP_2018,Furubayashi_JNM_2009,Uytven_NF_2022}.


Achieving fusion necessitates heating plasma to staggering temperatures, far beyond hundreds of millions of degrees, and maintaining this high-density plasma for an extended period. This demanding criterion places extraordinary stress on the materials utilized in the plasma-facing components and the divertor target plates of fusion devices, given the immense heat and plasma flux that emanate from the device. Neutral particles are pivotal in mitigating the particle and energy flux directed towards the divertor target \cite{Dekeyser2014}. Under these conditions, a neutral buffer layer emerges in front of the divertor target, which enhances ion-neutral interactions and shields the target from the direct onslaught of high-energy plasma. Consequently, the accurate simulation of the plasma edge, encompassing both plasma and neutral particles, is indispensable for devising operational strategies and blueprinting the next generation of fusion reactors \cite{Hoshino_cpp_2008,Karney_cpp_1998}.

In the edge plasma, the plasma can be described by macroscopic fluid equations. For the neutral particles, regions with high collision rates approach the hydrodynamic limit and can be effectively described using macroscopic fluid dynamics equations. Conversely, regions with low collision rates, being far from thermal equilibrium, require kinetic equations for precise characterization \cite{Maes_POP_2023,Riemann_CPP_2004,Rensink_CPP_1998}. This presents a challenge: relying solely on macroscopic fluid equations falls short in accurately depicting the low-collision regions, while an exclusive dependence on kinetic equations is computationally intensive and resource-demanding due to their inherent high dimensionality \cite{Valentinuzzi_CPP_2018,Horsten_CPP_2016}.

Traditionally, the direct simulation Monte Carlo (DSMC) method has been employed for simulating rarefied gas dynamics. However, this approach can result in excessively long computation times and large consumption of computer memory due to the higher frequency of charge exchange collisions. That is, the time step and spatial cell size should be smaller than the mean collision time and mean free path of the neutral particles, respectively. Moreover, the inherent statistical noise in DSMC can impede convergence when dealing with neutral particle models integrated with plasma fluid equations. Consequently, it is imperative to develop an efficient and accurate computational method that possesses asymptotic preserving properties (enabling the use of fewer spatial cells) and rapid convergence characteristics (allowing for a reduced number of iteration steps). Such a method is essential for the design of next-generation nuclear fusion reactors, as indicated by recent studies \cite{Uytven_CPP_2020,Blommaert_NME_2019}.

As detailed in the review article~\cite{DSA2002}, a significant breakthrough in the efficient simulation of kinetic equations emerged in the field of neutron transport, where the kinetic equation is solved concurrently with the synthetic equation of the diffusion type. This coupling strategy has been successfully applied to specific rarefied gas flow problems, such as planar Poiseuille flow~\cite{Valougeorgis:2003zr} where the flow direction is perpendicular to the computational domain, and has ultimately evolved into the general synthetic iteration scheme (GSIS) for general rarefied gas flows~\cite{SuArXiv2019}. GSIS effectively spans the micro-macro gap by iteratively solving the kinetic equation and its corresponding synthetic equations. Within the synthetic equation, the linear Navier-Stokes-Fourier (NSF) equations are integrated with higher-order terms (HoTs), which are directly derived from the velocity distribution function. These HoTs capture rarefaction effects and ensure the validity of the constitutive relations across all Knudsen numbers. In the continuous flow regime, where the Knudsen number is low, the higher-order terms diminish to zero, ensuring that GSIS asymptotically preserves the Navier-Stokes equations in the continuum limit. This preservation allows for the use of larger cell sizes and time steps, which is particularly beneficial for simulating edge plasma.
Through rigorous mathematical analysis and extensive numerical testing~\cite{SU2020SIAM, zeng2023general, liuw_JCP2024,yanbin_CF2024}, GSIS has demonstrated a significant reduction in the number of iterations required and an enhancement in simulation efficiency by several orders of magnitude.



This study will focus on developing a new GSIS algorithm for neutral particles in the plasma edge, aiming to achieve rapid convergence and asymptotic preservation. This approach is intended to reduce the number of spatial cells (simulation memory) and iterations (simulation time) required. The structure of this paper is as follows. In Section~\ref{model} we introduce the kinetic model equations and hydrodynamic equations for neutral particles. In Section~\ref{sec:GSIS}, we introduce the detailed numerical method to solve the GSIS. In Section~\ref{Sec:num_example}, the effectiveness of the GSIS algorithm is validated in various numerical simulations. Finally, the conclusions are given in Section~\ref{Sec:conclusion}.

\section{Flow control equations }\label{model}

In the field of edge plasma research, the dynamics of plasma are generally characterized by macroscopic fluid equations, whereas neutral particles are modeled using the Boltzmann kinetic equation. Given the high-dimensional complexity of the Boltzmann equation, which leads to substantial computational challenges, the computational speed in coupled simulations of plasma and neutral particle flows is predominantly limited by the solutions for the neutral particles. This study concentrates on accelerating the resolution of the Boltzmann equation, rather than addressing the dynamic evolution of both plasma and neutral particles concurrently. Consequently, we examine neutral particles within a static plasma background, which allows for a more focused development and assessment of the GSIS in the context of edge plasma issues.

\subsection{Kinetic model equation }

The kinetic equation for neutral gas reads \cite{HorstenNF2017,StaceyNF2017} 
\begin{equation} \label{eq:kinetic}
    \begin{aligned}
         \frac{\partial {f}}{\partial t}  + \bm{\xi}\cdot \nabla {f}= n_e K_r f_D  - n_e K_D {f} + k_{cx}  ({n}f_D -n_D{f}), 
    \end{aligned}
\end{equation}
where ${f}(t,\bm{\xi},\bm{x})$ is the neutral distribution function, $t$ is the time, $\bm{\xi}$ is the neutral particle velocity, and $\bm{x}$ is the spatial coordinates. Macroscopic properties can be obtained by taking the moments of the velocity distribution function, i.e., the number density $n$, bulk velocity $\bm{u}$, temperature $T$, stress tensor $\Pi$, and heat flux $\bm{q}$ of the neutral gas can be calculated as
\begin{equation}\label{eq:Macroscopic}
    \left[n,n\bm{u},\frac{3}{2}nk_BT,\bm{\Pi},\bm{q}\right]= \int \left[1,\bm{\xi},\frac{m}{2}c^2,m\left(\bm{cc}-\frac{c^2}{3}\bm{I}\right),\frac{m}{2}c^2\bm{c} \right] fd\bm{\xi},
\end{equation}
where $\bm{c}=\bm{v}-\bm{u}$ is the peculiar velocity, $\bm{I}$ is a $3 \times 3$ identity matrix. Here, an integral without bounds is used to denote the integral over the three-dimensional velocity space. The gas pressure is $p = n k_BT $, with $k_B$ being the Boltzmann constant. 

\begin{figure}[t]
    \centering
    \includegraphics[width=0.5\linewidth]{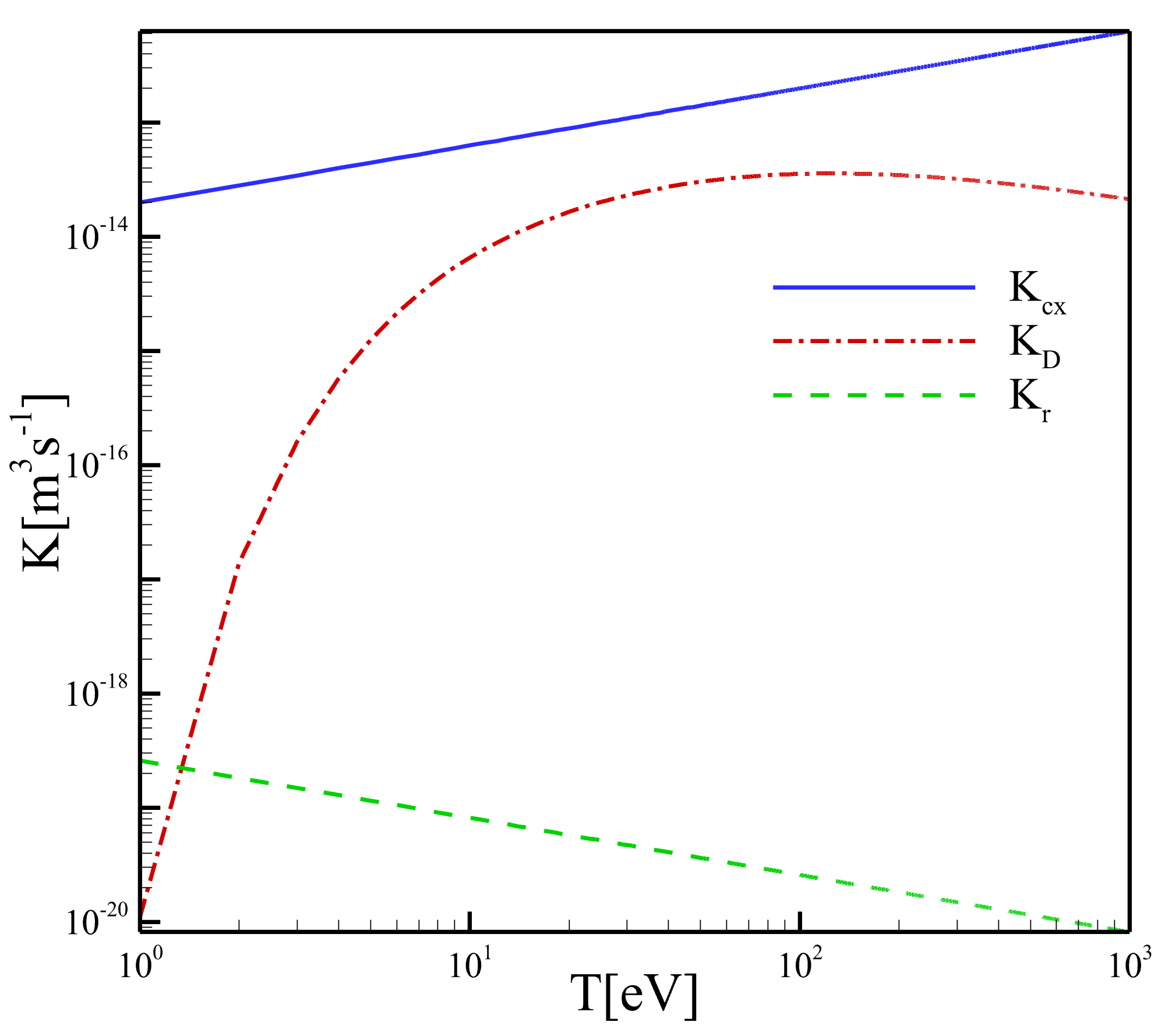}
    \caption{The coefficients of radiative recombination, ionization, and charge exchange as per Eq.~\eqref{eq:K}.}
    \label{fig:K}
\end{figure}

Equation~\eqref{eq:kinetic} contains three types of collisions: the radiative recombination with the rate coefficient $K_r$, the electron impact ionization with the rate coefficient $K_D$, and the charge-exchange with the rate coefficient $K_{cx}$. The expressions for the rate coefficients are adopted from Ref.~\cite{Dekeyser2014}:
\begin{equation} \label{eq:K}
    \begin{aligned}
        K_r &= 0.7\times10^{-19}\left(\frac{13.6e}{k_B T_e}\right)^{\frac{1}{2}},\\
        K_D &= \frac{2\times 10^{-13}}{6+\frac{k_B T_e}{13.6e}}
        \left(\frac{k_B T_e}{13.6e}\right)^{\frac{1}{2}}
        \exp\left(-\frac{13.6e}{k_B T_e}\right),\\
        {K_{cx}} &= 3.2\times 10^{-15}\left(\frac{k_B T_D}{0.026e}\right)^{\frac{1}{2}},
    \end{aligned}
\end{equation}
where $e$ is the elementary charge, $T_e$ and $T_D$ are the electron and ion temperatures, respectively. The three coefficients are shown in Fig.~\ref{fig:K} over a temperature range. 

The ion and electron densities are $n_D$ and $n_e =Z_D n_D$, respectively.  For pure Deuterium, we have $Z_D = 1$. The ion distribution is assumed to follow a Maxwell-Boltzmann distribution with ion temperature \(T_D\) and ion macroscopic velocity $\bm{u_D}$:
\begin{equation}\label{eq:Maxwell-distribution}
    f_D = n_D \left( \frac{m}{2 \pi k_B T_D} \right)^{3/2}
    \exp\left( - \frac{m(\bm{\xi} - \bm{u}_D)^2}{2 k_B T_D} \right).
\end{equation}

Typically, the dimensionless Knudsen number, defined as $\text{Kn} = \lambda/L$, is used to assess the fluid behavior, where $\lambda$ is the mean-free path and $L$ is a characteristic flow length. A smaller Knudsen number indicates stronger fluid-like behavior. Here we take $L$ to be the local temperature decay length, given by $L = T/||\nabla T||_2$. By considering only the mean-free path due to charge-exchange collisions, the Knudsen number can be calculated as
\begin{equation} \label{eq:kn}
\text{Kn} = \frac{||\nabla T||_2}{K_{cx} n_D T} \sqrt{\frac{3k_B T}{m}}.
\end{equation}

\subsection{Macroscopic equations}

By taking the moments of Eq.~\eqref{eq:kinetic}, the following macroscopic equations in accordance with the conservation of mass, momentum and total energy can be obtained:
\begin{equation} \label{eq:NS}
\begin{aligned}
   \partial_t n +\nabla \cdot ({n} {\bm{u}}) &= {S},\\
     \partial_t (nm\bm{u} )+ \nabla \cdot (m {n} \bm{u} \bm{u} + {p} \bm{I}+\bm{\Pi}_\text{NSF} ) &= \bm{S}_{mu},\\ 
      \partial_t (ne) +\nabla \cdot 
     \left[{n}e \bm{u} +(p\bm{I}+\bm{\Pi}_\text{NSF}) \cdot \bm{u} +\bm{q}_\text{NSF}\right]&= {S_E},
\end{aligned}
\end{equation}
where $e = \frac{3 }{2}k_BT + \frac{m}{2} u^2$;  ${S}$, ${\bm{S}_{mu}}$, and ${S_E}$ is respectively the source term for the creation and annihilation of neutral particle, momentum, and energy, given by
\begin{equation}
\begin{aligned}
    {S} &= n_D n_e K_r - {n} n_e K_D ,\\
     \bm{S}_{mu} &= m(n_D n_e K_r + {n} n_D K_{cx}) \bm{u_D} - m({n} n_e K_D + {n} n_D K_{cx}) \bm{u},\\ 
       {S_E} &= (n_D n_e K_r + {n} n_D K_{cx})\left(\frac{3}{2} k_B T_D  + \frac{m}{2} \bm{u}_D^2\right) 
  -({n} n_e K_D + {n} n_D K_{cx})\left(\frac{3}{2} k_B T_D + \frac{m}{2} \bm{u}^2\right).
\end{aligned}
\end{equation}

Equation~\eqref{eq:NS} is not closed due to the unknown expressions for the stress $\bm{\Pi}$ and heat flux $\bm{q}$. In the continuum limit, the Chapman-Enskog expansion of the kinetic equation provides the linear constitutive relations \cite{chapman1990mathematical}:
\begin{equation}\label{eq:HoT1}
\begin{aligned}
     \bm{\Pi}_{\text{NSF}} &= -\mu \left[\nabla \bm{u} + (\nabla \bm{u})^T - \frac{2}{3}(\nabla \cdot \bm{u})\bm{I}\right],\\
    \bm{q}_{\text{NSF}} &= -\kappa \nabla {T},
\end{aligned}
\end{equation}
where the viscosity $\mu$ and thermal conductivity $\kappa$ are
\begin{equation} \label{eq:viscosity}
\begin{aligned}
  \mu = \frac{{n} {k_B T}}{n_D k_{cx}}, \quad
  \kappa = \frac{5 {n} {k_B T}}{2m n_D k_{cx}}.
\end{aligned}
\end{equation}

However, these linear constitutive relations are inaccurate in highly non-equilibrium gas flow. The exact shear stress and heat flux without any truncation should be calculated directly from the velocity distribution functions as per Eq.~\eqref{eq:Macroscopic}, thus the solutions of macroscopic synthetic equations will be consistent with those of kinetic equations. This can only be extracted from the numerical solution of the kinetic equation \eqref{eq:kinetic}.


\section{The numerical methods}\label{sec:GSIS}

In this section, we describe the numerical methods for solving the mesoscopic kinetic equation and macroscopic synthetic equation in the finite volume framework, in an implicit manner.

\subsection{Conventional iterative scheme}

We first introduce the conventional iterative scheme (CIS), which is effective and accurate in finding steady-state solutions in transition and free-molecular flow regimes. However, for near-continuum flow, it converges very slowly and the converged solutions are usually wrong due to high numerical dissipation.  


The cell-centered scheme is adopted for spatial discretization. We use the subscripts $i,~j$ as the indices of the control cells, and let the subscript $ij$ denote the interface between adjacent cells $i$ and $j$. Then, for a given numerical time increment $\Delta t = t^{k+1} - t^{k}$ from step $k$ to $k+1$ in an implicit solving process, the discretized form of mesoscopic governing Eq.~\eqref{eq:kinetic} for neutral particle can be written as:
\begin{equation}\label{eq:11}
        \frac{f_{i}^{k+1}-f_{i}^k}{\Delta t} + \frac{1}{V_i}\sum_{j\in N(i)} \bm{\xi}_n f_{ij}^{k+1}S_{ij}= G_i^n-f_{i}^{k+1} B_i^k, \\
\end{equation}
where $S_{ij}$ and $V_i$ represent the area of face $ij$ and the volume of cell $i$, respectively; ${\xi}=\bm{\xi}\cdot \bm{n}$ is the molecular velocity component along normal direction pointing from cell $i$ to cell $j$; the sum of fluxes ${\bm{\xi_n}} f_{ij}$ is taken over all the faces of a cell $N(i)$. The sources terms in the right hand side of the above equation are 
\begin{equation}
    \begin{aligned}
        G_i &= f_{D,i}^k (n_{e,i}^k k_{r,i}^k + n_{i}^k k_{cx,i}^k ),\quad
        B_n^i = n_{e,i}^k k_{D,i}^k + n_{D,i}^k k_{cx,i}^k.
     \end{aligned}
\end{equation}

To apply a simple matrix-free implicit solving of the discretized equations, the incremental variable $\Delta f_i^k = f_i^{k+1}-f_i^{k}$ is introduced. And the delta-form discretized kinetic equation $\Delta f_i^k$ is given by:
\begin{equation}
        \left(\frac{1}{\Delta t}+B_i^k\right)\Delta f_i^k + \frac{1}{V_i}\sum_{j\in N(i)} \xi_n \Delta f_{ij}^{k}S_{ij}= G_i^k-f_{i}^{k} B_i^k -\frac{1}{V_i}\sum_{j\in N(i)} \xi_n f_{ij}^{k}S_{ij}.
        \label{eq:delta_form}
\end{equation}

The interface fluxes $f^k_{ij}$ in the right-hand-side of Eq.~\eqref{eq:delta_form} are reconstructed using a second order upwind scheme. Specifically, we have $f_{ij}^k=[\xi_n^+(f_i+\phi \nabla f_i\cdot \bm{x})+\xi_n^-(f_j + \phi \nabla f_j\cdot \bm{x})]/2$ where $\xi_n^{\pm}=[1\pm \text{sign}(\xi_n)]/2$ denotes the interface sign directions with respect to the cell center value, and $\phi$ is the Venkatakrishnan limiter. the derivative information is obtained via the least squares method. The increment fluxes $\Delta f_{ij}^k$ in the right hand side of Eq.~\eqref{eq:delta_form} are constructed using a first-order upwind scheme, given by $\Delta f_{ij}^k=[\xi_n^+ \Delta f_i +\xi_n^- \Delta f_j]/2$ This is because the discretization accuracy of the left-hand-side of the equation does not affect the
accuracy of the equation after convergence. Finally, Eq.~\eqref{eq:delta_form} can be rewritten as:
\begin{equation}\label{eq:17}
        d_i \Delta f_i^k + \sum_{j \in N(i)} d_j \Delta f_j^k = r_i^k,  
\end{equation}
where the matrix elements are
\begin{equation}\label{eq:18}
    \begin{aligned}
        d_i &= \frac{1}{\Delta t}+B_i^k + \frac{1}{2V_i}\sum_{j\in N(i)} \xi_n \xi^+_n S_{ij}, \\
        d_j &=\frac{1}{2V_i}\sum_{j\in N(i)} \xi_n \xi^-_n S_{ij},\\
        r_i^k &= G_i^k-f_{i}^{k} B_i^k -\frac{1}{2V_i}\sum_{j\in N(i)} \xi_n (\xi^+_n f_i^k +  \xi^-_n f_j^k)S_{ij}.
     \end{aligned}   
\end{equation}

The above matrix-free equations can be solved using standard Lower-Upper Symmetric Gauss-Seidel (LU-SGS) technology. At steady state, the increment $\Delta f_i^k$ is zero, indicating that the mesoscopic residuals $r_i^k$ also converge to zero. Here, $d_i$ and $d_j$ represent the main diagonal and non-diagonal elements of the matrix, respectively. In structured grids, the matrix has the property of main diagonal dominance, which improves the efficiency of the LU-SGS process.

 
\subsection{The GSIS}

To accelerate convergence and reduce numerical dissipation in near-continuum flow regime, GSIS has recently been proposed for dilute gas \cite{SuArXiv2019, zeng2023general, liuw_JCP2024,yanbin_CF2024}. The key idea of GSIS is to introduce the macroscopic synthetic equations to guide the evolution of the macroscopic quantities appear in the collision operator and the corresponding velocity distribution function. To be specific, from the distribution function at the $k$-th iteration step, we first find the distribution function at the intermediate (i.e., $k+1/2$) iteration step using the CIS:
\begin{equation}\label{eq:11}
        \frac{f_{i}^{k+1/2}-f_{i}^k}{\Delta t} + \frac{1}{V_i}\sum_{j\in N(i)} \xi_n f_{ij}^{k+1/2}S_{ij}= G_i^n-f_{i}^{k+1/2} B_i^k.
\end{equation}
Note that the superscript $k+1/2$ means that the distribution function solved at this stage has not been modified by the solutions of synthetic equations. 


Then, we use the solution of macroscopic synthetic equation
\begin{equation} \label{eq:NS_Correct}
\begin{aligned}
     \partial_t n +\nabla \cdot ({n} {\bm{u}}) &= {S},\\
     \partial_t (nm\bm{u}) +\nabla \cdot (m {n} \bm{u} \bm{u} + {p} \bm{I}+\bm{\Pi}_\text{NSF} ) &= \bm{S}_{mu} - \nabla \cdot \bm{\text{HoT}_{\Pi}},\\ 
     \partial_t (ne) + \nabla \cdot 
     \left[{n}e \bm{u} +(p\bm{I}+\bm{\Pi}_\text{NSF}) \cdot \bm{u} +\bm{q}_\text{NSF}\right]&= {S_E}- \nabla \cdot (\bm{\text{HoT}_{q}} + \bm{u}\bm{\text{HoT}_{\Pi}}),
\end{aligned}
\end{equation}
to guide the evolution of the distribution function towards the steady state. It is important to note that only the HoTs are extracted from $f_{k+1/2}$:
\begin{equation}
        \begin{aligned}
        \text{HoT}_{\bm{\Pi}} &= \int m\left(\bm{c}\bm{c}-\frac{c^2}{3}\bm{I}\right)f^{k+1/2} d\bm{v} -\bm{\Pi}_{\text{NSF}}^{k+1/2},\\
        \text{HoT}_{\bm{q}} &= \int \frac{m}{2}c^2\bm{c}f^{k+1/2} \mathrm{d}\bm{v} -\bm{q}_{\text{NSF}}^{k+1/2}.
        \end{aligned}
        \label{eq:14}
\end{equation} 

It is important to highlight that the macroscopic quantities in the left-hand side of Eq.~\eqref{eq:NS_Correct} is solved during the $(k+1)$-th iteration step, whereas the terms on the right-hand side are resolved in the intermediate iteration step. Therefore, they cannot cancel each other out until the convergence is achieved. As a result of this arrangement, fast convergence and asymptotic preserving are achieved \cite{SU2020SIAM, liuw_JCP2024}. When the steady state is reached, Eq.~\eqref{eq:NS_Correct} exactly reduces to Eq.~\eqref{eq:NS}, which is exactly derived from the kinetic equaiton \eqref{eq:kinetic}.

We use the following finite volume method to solve the macroscopic synthetic equations \eqref{eq:NS_Correct}:
\begin{equation}
        \frac{\bm{W}^{k+1}_i-\bm{W}^K_D}{\Delta t} + \frac{1}{V_i} \sum_{j\in N(i)} \left[\bm{F}_{c,ij}^{k+1}+\bm{F}_{v,ij}^{\text{NSF},k+1} \right]S_{ij}=\bm{Q}_i^{k+1}- \frac{1}{V_i}\sum_{j\in N(i)} \bm{F}_{v,ij}^\text{HoT}S_{ij},
\end{equation}
where the macroscopic variables $\bm{W}$ and the fluxes $\bm{F}$ (including the convective flux $\bm{F_c}$ and viscous flux $\bm{F_v}$), in two-dimensional space, are 
\begin{equation}
\mathbf{W}=\left[\begin{array}{c}
n \\
nm u_{x} \\
nm u_{y} \\
n e
\end{array}\right], \quad \mathbf{F}_{\mathrm{c}}=\left[\begin{array}{c}
n u_{n} \\
nm u_{x} u_{n}+n_{x} p \\
nm u_{y} u_{n}+n_{y} p \\
u_{n}(nm e+p)
\end{array}\right], \quad \mathbf{F}_{\mathrm{v}}(\Pi, \mathbf{q})=\left[\begin{array}{c}
0 \\
n_{x} \Pi_{x x}+n_{y} \Pi_{x y} \\
n_{x} \Pi_{y x}+n_{y} \Pi_{y y} \\
n_{x} \Theta_{x}+n_{y} \Theta_{y}
\end{array}\right],
\end{equation}
with $\Theta_x = u_x \Pi_{xx}+u_y \Pi_{xy} +q_x$ and $\Theta_y = u_x \Pi_{yx}+u_y \Pi_{yy} +q_y$. Here, $u_n=u_x n_x +u_y n_y $ is defined as the scalar product of the macro velocity vector and the unit normal vector of the cell face. $\bm{F}_{v,ij}^{\text{NSF},k+1}=\bm{F_v}(\Pi_\text{NSF},q_\text{NSF})$, $ \bm{F}_{v,ij}^{\text{HoT},k+1}=\bm{F_v}(\text{HoT}_{\Pi},\text{HoT}_{q})$, and the source terms $\bm{Q}$ are 
\begin{equation}
\mathbf{Q}=\left[\begin{array}{c}
S \\
S_{mu_x} \\
S_{mu_y} \\
 {S_E}
\end{array}\right]. 
\end{equation}

Introducing the incremental variables $\Delta \bm{W^m_i} = \bm{W}^{m+1}_i-\bm{W}^m_i$  with $m$ being the inner iteration index in solving macroscopic equations, the delta-form discretized synthetic equations become: 
\begin{equation}\label{eq:delta_Macroscopic_equations}
    \begin{aligned}[b]
        \left[\frac{1}{\Delta t_i}- \left(\frac{\partial \bm{Q}}{\partial \bm{W}}\right)^{m}\right]\Delta \bm{W}_i^{m} + \frac{1}{V_i} \sum_{j\in N(i)} \Delta\bm{F}_{ij}^{m}S_{ij}=\underbrace{-\frac{1}{V_i}\sum_{j\in N(i)} \bm{F}_{ij}^{m}S_{ij}+\bm{Q}_i^{m}}_{R_i^m},
    \end{aligned}
\end{equation}

The general form of the macroscopic fluxes can be expressed as $F_{ij} = F(W_L, W_R, S_{ij})$, where $W_{L,R}$ represents the reconstructed values of the left and right sides of the interface, respectively, and can be further written as $W_{L/R}=W_{i/j}+\phi \nabla (W_{i/j}\cdot \bm{x})$,with $\phi$ being the Venkatakrishnan limiter. For the reconstruction of the macroscopic flux, the Rusanov scheme~\cite{mohamed2021modified} is applied, while the gradient and the limiter are chosen to be consistent with the mesoscopic equations. 

To obtain a matrix-free form, the implicit fluxes in the macroscopic system Eq.~\eqref{eq:delta_Macroscopic_equations} are approximated by the Euler-type fluxes:
\begin{equation}\label{eq:macro_euler}
\begin{aligned}
    \Delta \bm{F}_{ij}^m &= \frac{1}{2}\left[\Delta \bm{F}_i^m + \Delta \bm{F}_j^m+\Gamma_{ij}(\Delta\bm{W}_i^m - \Delta\bm{W}_j^m)\right], \\  
    \Gamma_{ij} &= |u_n| + c_s + \frac{2\mu}{nm|\bm{n}_{ij}\cdot(\bm{x}_j-\bm{x}_i)|}.
    \end{aligned}
\end{equation}
Since the control volume satisfies the geometric conservation law, the interface fluxes through the cell accumulate to  $\sum_{j\in N(i)}\bm{F}_{i}S_{ij}=0$. While the flux can be directly represented by the convective flux $\bm{F}=\bm{F}_c$, the flux of subscript $j$ can be written as a matrix free form $\Delta \bm{F}_j^m=\bm{F}(\bm{W}_j^m + \Delta \bm{W}_j^m) - \bm{F}(\bm{W}_j^m)$. 
Substituting Eq.~\eqref{eq:macro_euler} into Eq.~\eqref{eq:delta_Macroscopic_equations}, the implicit governing equations for macroscopic variables become:
\begin{equation}\label{eq:res_macro}
    D_i\Delta \bm{W}_i^{m} + \frac{1}{2V_i}\sum_{j\in N(i)} \left(\Delta\bm{F}_{j}^{m} - \Gamma_{ij}\Delta \bm{W}_j^m\right)S_{ij}=\bm{R}_i^m,
\end{equation}
where
\begin{equation}
    D_i = \frac{1}{\Delta t_i}+\frac{1}{2V_i}\sum_{j\in N(i)}\Gamma_{ij}S_{ij}- \left(\frac{\partial \bm{Q}}{\partial \bm{W}}\right)^m_i.
\end{equation}

The above equations can be solved by the standard LU-SGS technique.


\subsection{Overview of the GSIS}

\begin{figure}[t]
    \centering
    \includegraphics[width=0.8\linewidth]{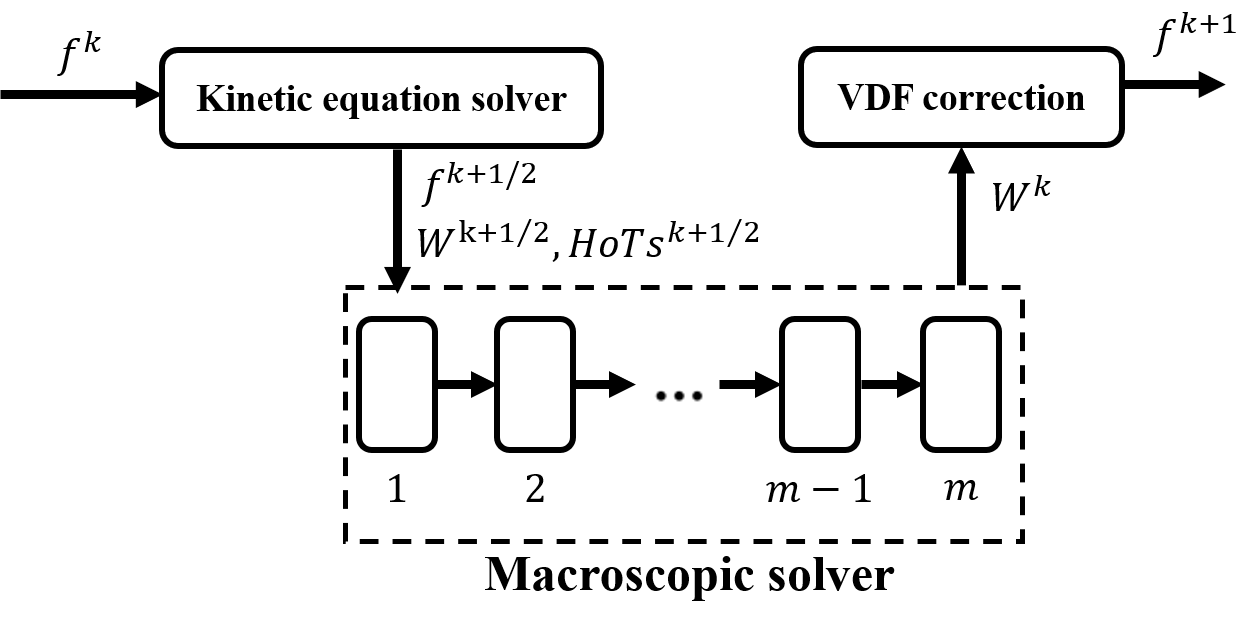}
    \caption{The flowchart of GSIS. The kinetic equation, which is solved by the CIS, provides HoTs to the macroscopic synthetic equations, which are essentially the NSF equations with source terms from the HoTs. Solving these equations yields a solution that is more close to the steady state. The obtained macroscopic quantities are then used to update the velocity distribution function via Eq.~\eqref{vdf_correction}, until the system reaches a steady state.
}
    \label{fig:GSIS}
\end{figure}

Here, we present the proposed GSIS algorithm for the kinetic equation of edge plasma flow. The overall computational process is a nested loop, with the outer and inner loop indices denoted as $k$ and $m$, as shown in Fig.~\ref{fig:GSIS}. The outer loop solves the kinetic equation~\eqref{eq:kinetic} by the CIS, while the inner loop solves the macroscopic equations using the LU-SGS technique. Each inner step starts from the current time step with the latest macroscopic state, together with the HoTs and boundary conditions from the current step in the outer loop. In summary, the procedures of GSIS can be described as follows:
\begin{enumerate}

    \item Calculate the numerical fluxes $f_{ij}^k$ using the distribution function $f^n$ and the equilibrium state $g^k$ by the macroscopic variables $W^k$.
    
    \item Evaluate the mesoscopic residual $r^k$ in Eq.~\eqref{eq:18}, and solve Eq.~\eqref{eq:17} to obtain $\Delta f^k$, then get the VDF from $f^{k+1/2} = f^k + \Delta f^n$. 
    
    \item Calculate the conservative variables $\bm{W}^{k+1/2}$ by taking moments of $f^{k+1/2}$ in Eq.~\eqref{eq:Macroscopic}, as well as HoTs  defined in Eq.~\eqref{eq:14} .

    \item Evaluate the macroscopic residual $R^{m}$ in Eq.~\eqref{eq:res_macro} using the macroscopic variables and boundary fluxes obtained in Step 3. Solve Eq.~\eqref{eq:delta_Macroscopic_equations} to obtain $\Delta \bm{W}^{m}$, repeated until criterion is reached or the maximum number of iteration is reached, then update the macroscopic variables $\bm{W}^{m+1}$.

    \item When the macroscopic conservative variables $\bm{W}^{k+1}$ are solved, they are used to update the velocity distribution function. That is, the non-equilibrium part is kept while the equilibrium is modified
    \begin{equation}\label{vdf_correction}
        f^{k+1}=f^{k+1/2}+[f_{eq}(\bm{W}^{k+1})-f_{eq}(\bm{W}^{k+1/2})]
    \end{equation}
    
    \item Repeat the above steps until convergence.
\end{enumerate}

\section{Numerical results}\label{Sec:num_example}

In this section, several edge plasma flows are simulated to assess the GSIS.  The convergence criterion for the kinetic equation is that the volume-weighted relative change of the moments (density, velocity, and total temperature) between two consecutive iterations 
\begin{equation}
E^k =  \frac{\sqrt{\sum_i(\phi_i^k-\phi_i^{k-1})^2 d\Omega}}{\sqrt{\sum_i(\phi_i^{k-1})^2 d\Omega}}|_{max} , \quad \phi = \in (\rho,\textbf{u},T),
\end{equation}
is smaller than $\epsilon = 10^{-5}$. For the macroscopic solver, the iteration process was terminated either when $\epsilon = 10^{-5}$ is reached or when the maximum number of iterations reached 1000. We will initially assess the GSIS with the parameters outlined in Eq.~\eqref{eq:K}, where the charge-exchange collision is dominant. Subsequently, we will increment $k_D$ by one and two orders of magnitude to determine if the favorable characteristics of GSIS remain intact.

\begin{figure}[h]
    \centering
    \vspace{0.5cm}
    \includegraphics[width=1\linewidth]{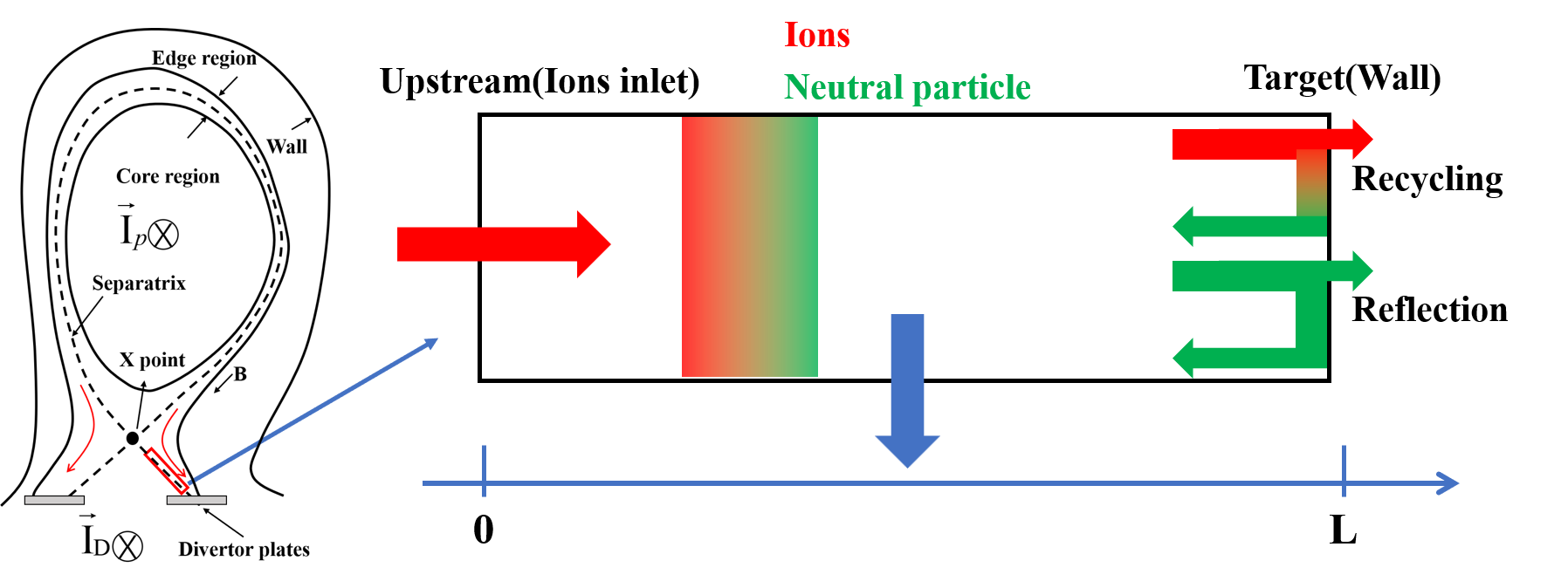}
    \caption{Sketch of the one-dimensional model, where the left side shows the schematic of the edge region in the poloidal plane of a divertor tokamak, with the red box indicating the simulation region, as shown in the right-side diagram. This region is then simplified into a one-dimensional model.
 }
    \label{fig:plasmaSetup}
\end{figure}

\subsection{One-dimensional edge plasma flow}


Figure~\ref{fig:plasmaSetup} provides a schematic of the transition from the two-dimensional poloidal plane to a one-dimensional model that only includes poloidal variations. The left figure shows the cross-section of the divertor, and the region within the red rectangle is selected as the computational domain. In the right figure, the red and green arrows represent the ion and neutral flows, respectively. At the target, a small fraction $R^i$ of the incident plasma flux density is recycled as neutral particles, and a small fraction $R^n$ of the incident neutral particle flux density is reflected. The red and green shaded areas in the background depict the ionization of the recycled and reflected neutral particles. 

At the upstream boundary, the boundary condition for neutral particles is a vacuum exit condition, so the incoming flux is zero, i.e., $f(\xi) = 0$ when $\xi > 0$. 
At the target position, a fraction $R^D$ of the incident ions is recycled as neutrals, and a fraction $R^n$ of the incident neutrals is reflected. The recycled or reflected neutral has a probability $R^T$ of being thermally released as a Franck-Condon dissociated neutral with an energy of 2 eV. The remaining fraction is emitted as fast neutrals that obtain a fraction $\alpha$ of the energy of the incident particle. Then, the target boundary condition is:
\begin{equation}
\begin{aligned}
f(\xi)=
&R^{T} \delta\left(|\xi|-\xi_{T}\right) \int_{0}^{+\infty} \left[R^{D} f_{D}\left(\xi'\right)+R^{n} f\left(\xi'\right)\right] d \xi' \\
&+\frac{\left(1-R^{T}\right)}{\alpha}\left[R^{D} f_{D}\left(\frac{\xi'}{\sqrt{\alpha}}\right)+R^{n} f\left(\frac{\xi'}{\sqrt{\alpha}}\right)\right], \quad \text{for} \quad \xi <0,
\end{aligned}
\end{equation}
where \(f_{D}(\xi)\) is the ion distribution at the boundary, $\xi' = -\xi$, \(\xi_T\) is the thermal velocity when the particle energy is 2 eV, and \(\delta\) is the Dirac delta function. 

The one-dimensional kinetic neutral equation reads:
\begin{equation}
\frac{\partial{f}}{\partial t}+\xi\frac{\partial{f}}{\partial x}=n_e K_r f_D  - n_e K_D {f} + k_{cx}  ({n}f_D -n_D{f}), 
\end{equation}
where
\begin{equation}
    f_D = n_D \left( \frac{m}{2 \pi k_B T_D} \right)^{1/2}
    \exp\left( - \frac{m(\xi - u_D)^2}{2 k_B T_D} \right).
\end{equation}

\begin{figure}[t]
     \centering
\subfloat{{\includegraphics[width=0.25\textwidth]{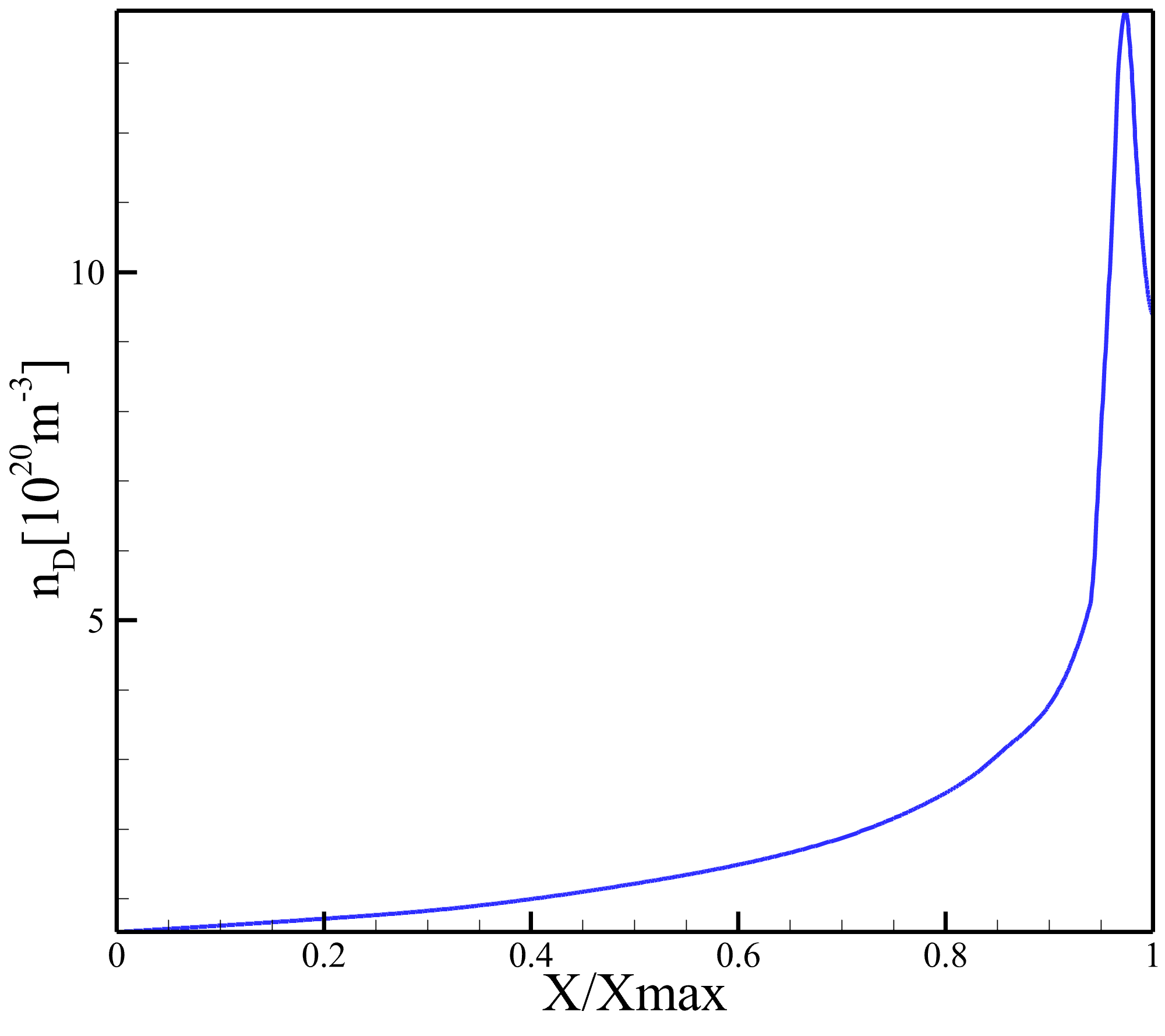}}}
\subfloat{{\includegraphics[width=0.25\textwidth]{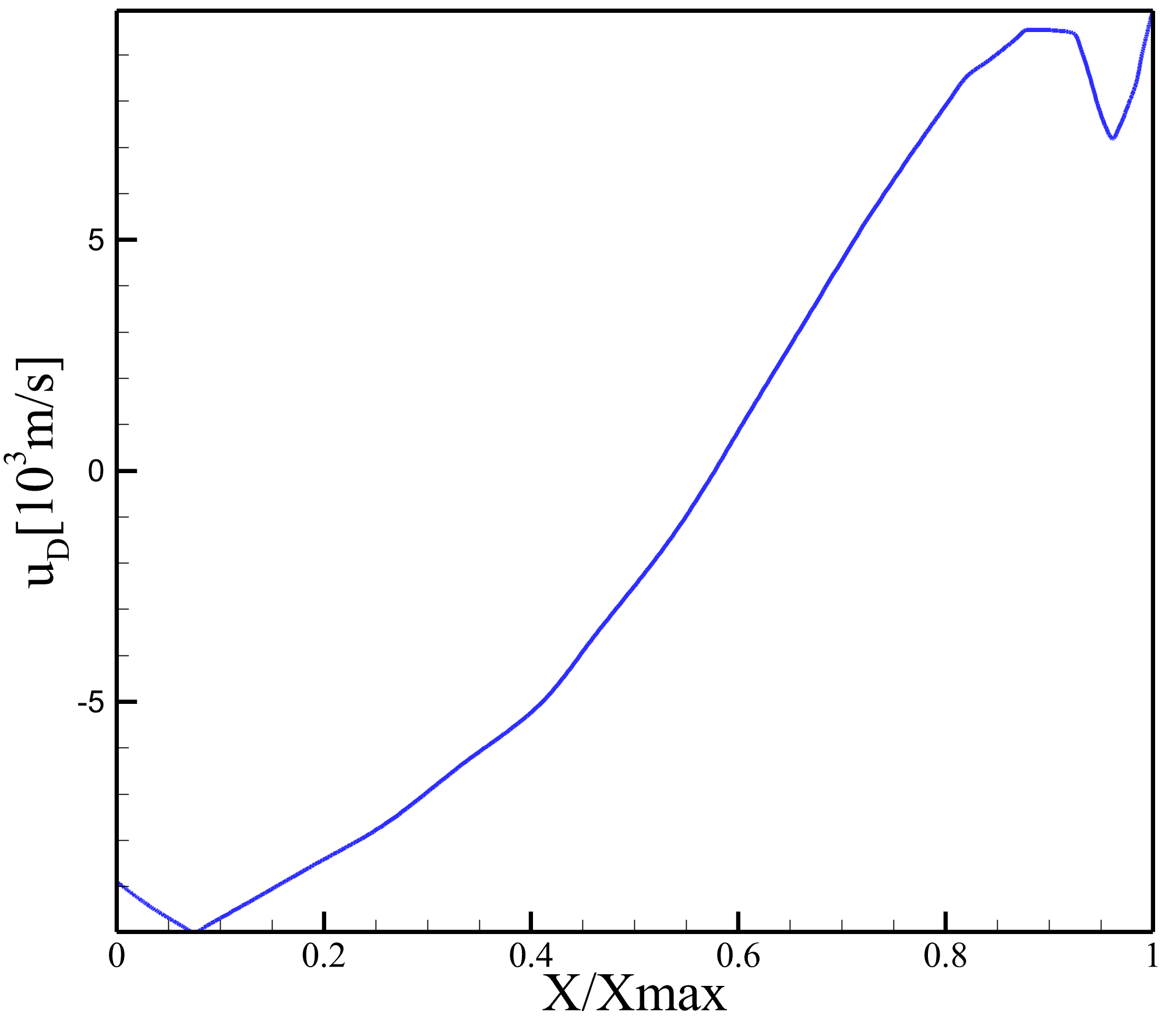}}}
\subfloat{{\includegraphics[width=0.25\textwidth]{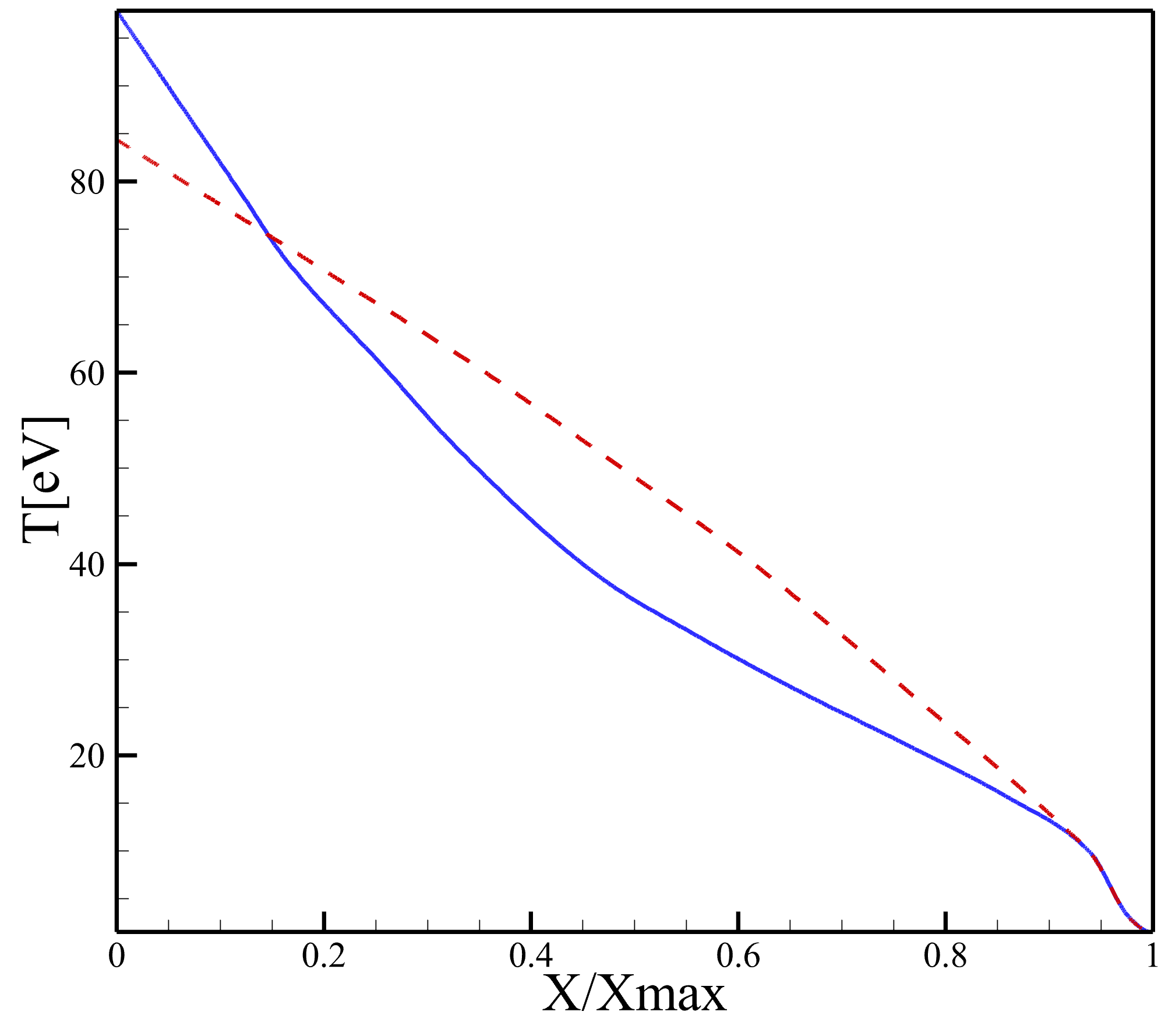}}}
\subfloat{{\includegraphics[width=0.25\textwidth]{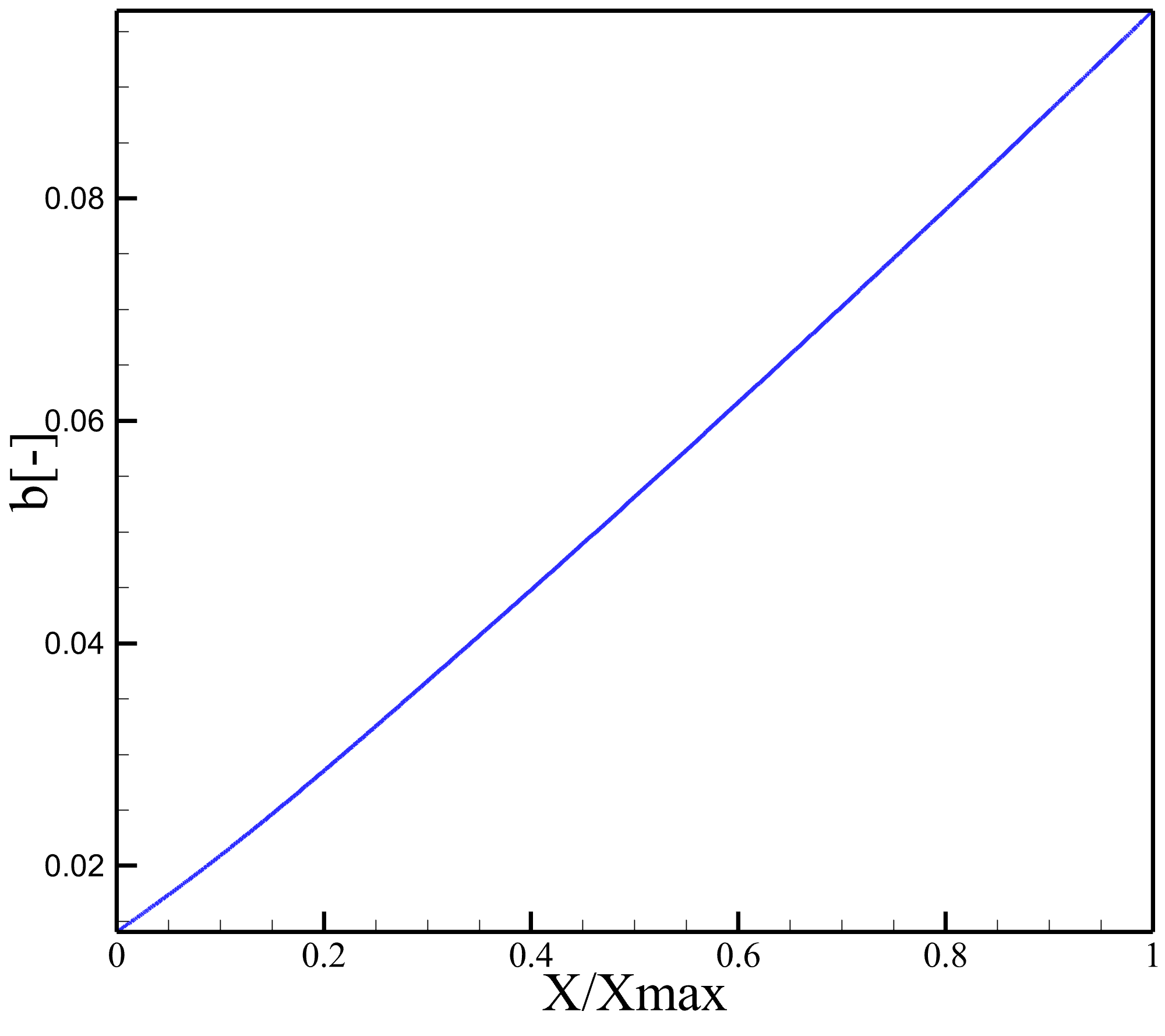}}}
\\
\subfloat{{\includegraphics[width=0.33\textwidth,  clip = true]{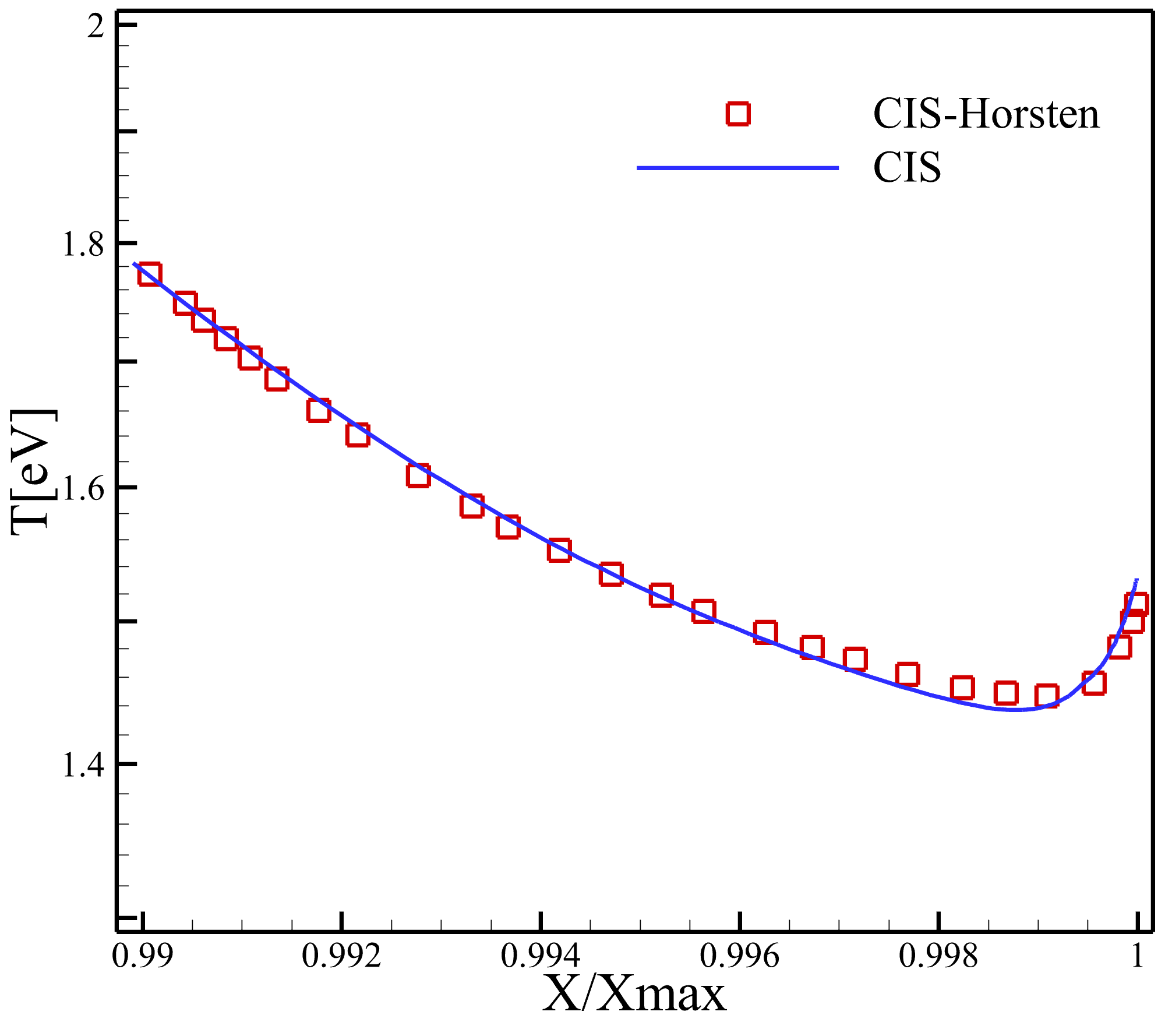}}}
\subfloat{{\includegraphics[width=0.33\textwidth,  clip = true]{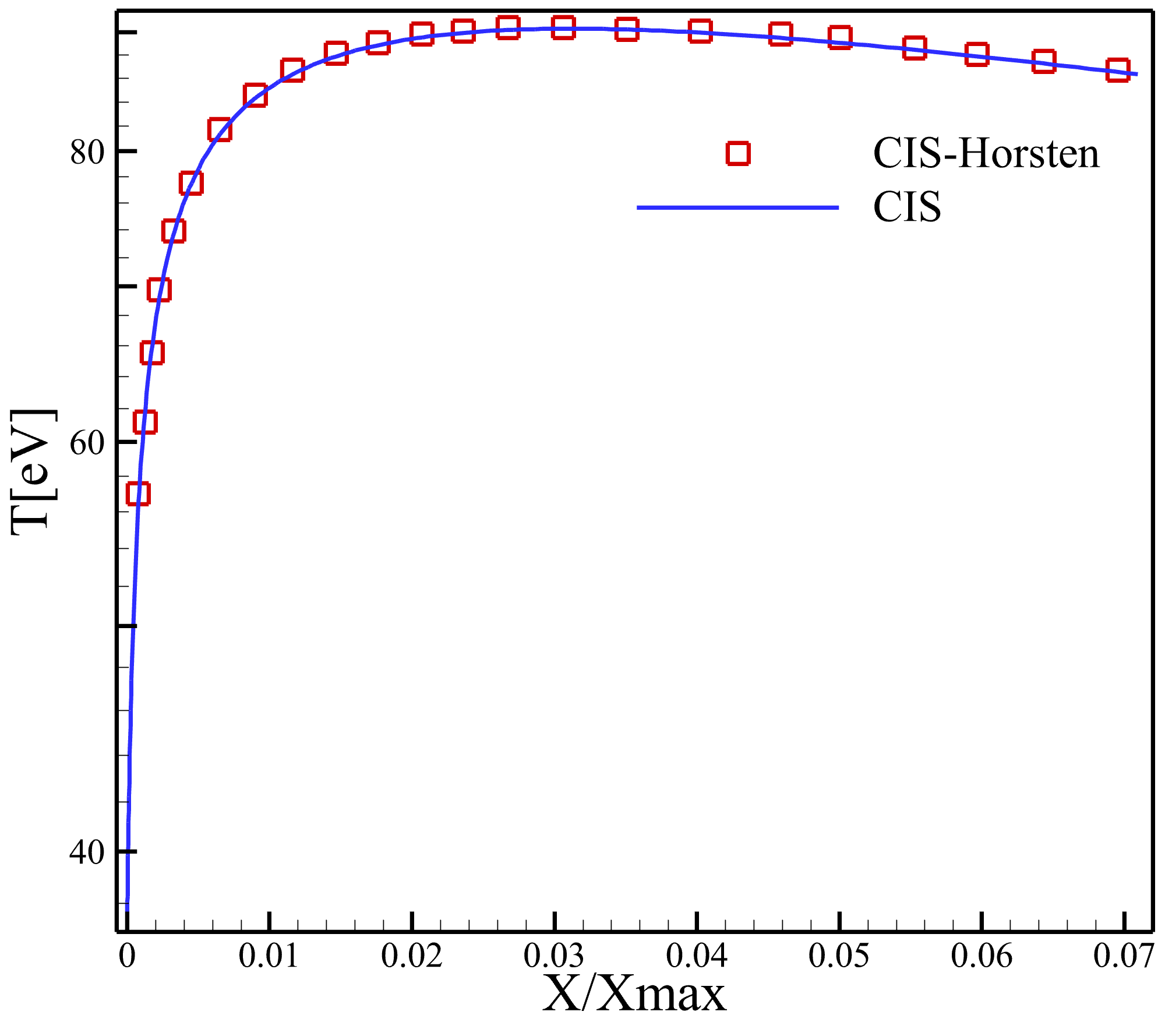}}}
\subfloat{{\includegraphics[width=0.33\textwidth,  clip = true]{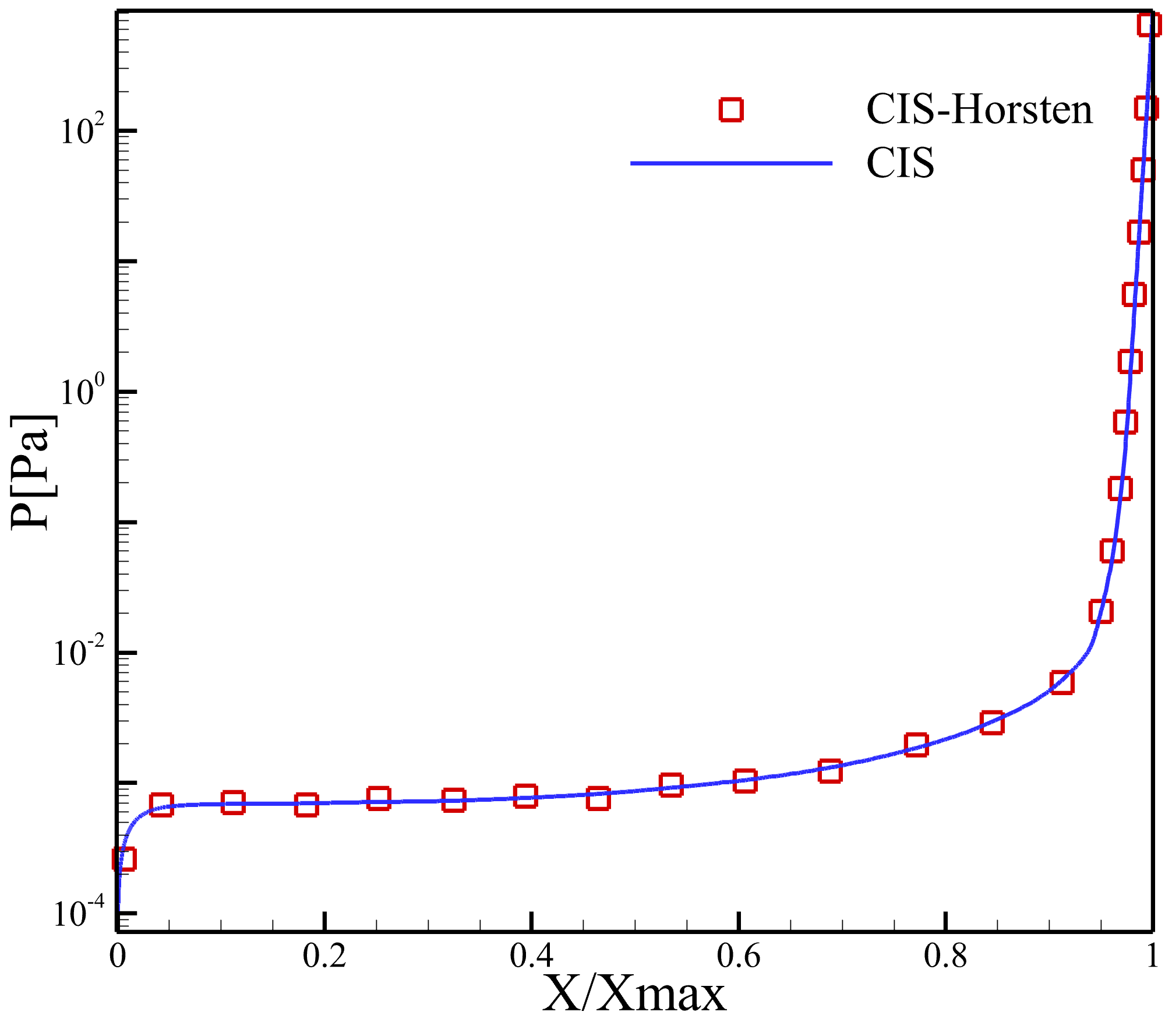}}}

 \caption{First row: the plasma background parameters. From left to right: ion density, ion velocity, electron (red dashed line) and ion (blue solid line) temperatures, and magnetic moment.  Second row: comparison between the the temperature near the upstream and downstream boundaries, as well as the global pressure, obtained from the CIS and 
Ref.~\cite{HorstenPOP2016}. }   
\label{fig:plasmaBG}
\end{figure}

The macroscopic equations are reduced to 
\begin{equation}
\begin{aligned}
      \partial_t n +\partial_x ({n} u_n) &= {S},\\
     \partial_t (nmu) +\partial_x (m {n} u_n^2 + {n}{k_B T}) &= S_{mu_n},\\
      \partial_t (ne) +\partial_x \left[\left(\frac{3}{2}{k_B T} + \frac{1}{2}mu^2\right)n u_n+ q\right] &= {S_E}, 
\end{aligned}
\end{equation}
where ${e}= \frac{1}{2}k_B T +\frac{1}{2}mu^2$, and 
\begin{equation}
\begin{aligned}
    {S} &= n_D n_e K_r - {n} n_e K_D,\\ 
    {S_{mu}} &= m(n_D n_e K_r + {n} n_D K_{cx}) u_D - m({n} n_e K_D + {n} n_D K_{cx}) u_n,\\
     {S_E} &= (n_D n_e K_r + {n} n_D K_{cx})\left(\frac{3}{2} k_B T_D + \frac{m}{2} u_D^2\right) 
  -({n} n_e K_D + {n} n_D K_{cx})\left(\frac{3}{2} k_B k_B  T_D + \frac{m}{2} u_n^2\right).
\end{aligned}
\end{equation}

Figure~\ref{fig:plasmaBG} illustrates the fixed background plasma state \cite{HorstenPOP2016}, In the test case, we choose $R^D=R^n=0.9, R^R=0.8,R^T=0.2$ and $\alpha=0.5$.
Due to the presence of numerical dissipation, solving the kinetic equation requires a refined spatial grid in CIS. Here we divide the X direction into $1000$ grid cells. The same grid is fully adequate for the fluid model, as the discretization error in the fluid model is sufficiently small even with a coarser grid compared to the grid requirements for solving the kinetic equation. The velocity space $\xi$ is divided into 200 grid cells, where the maximum velocity $10\sqrt{k_B T_{D,max}/m}$, with $T_{D,max}$ being the maximum ion temperature.

By comparing the temperature near the upstream and downstream boundaries and the global pressure  with the results from Ref.~\cite{HorstenPOP2016}, we find the CIS accurately solves the problem. Therefore, we will use the results obtained from the CIS as the reference solution for comparison with the GSIS results.

\begin{figure}[t]
     \centering
\subfloat[]{\includegraphics[width=0.4\textwidth, clip = true]{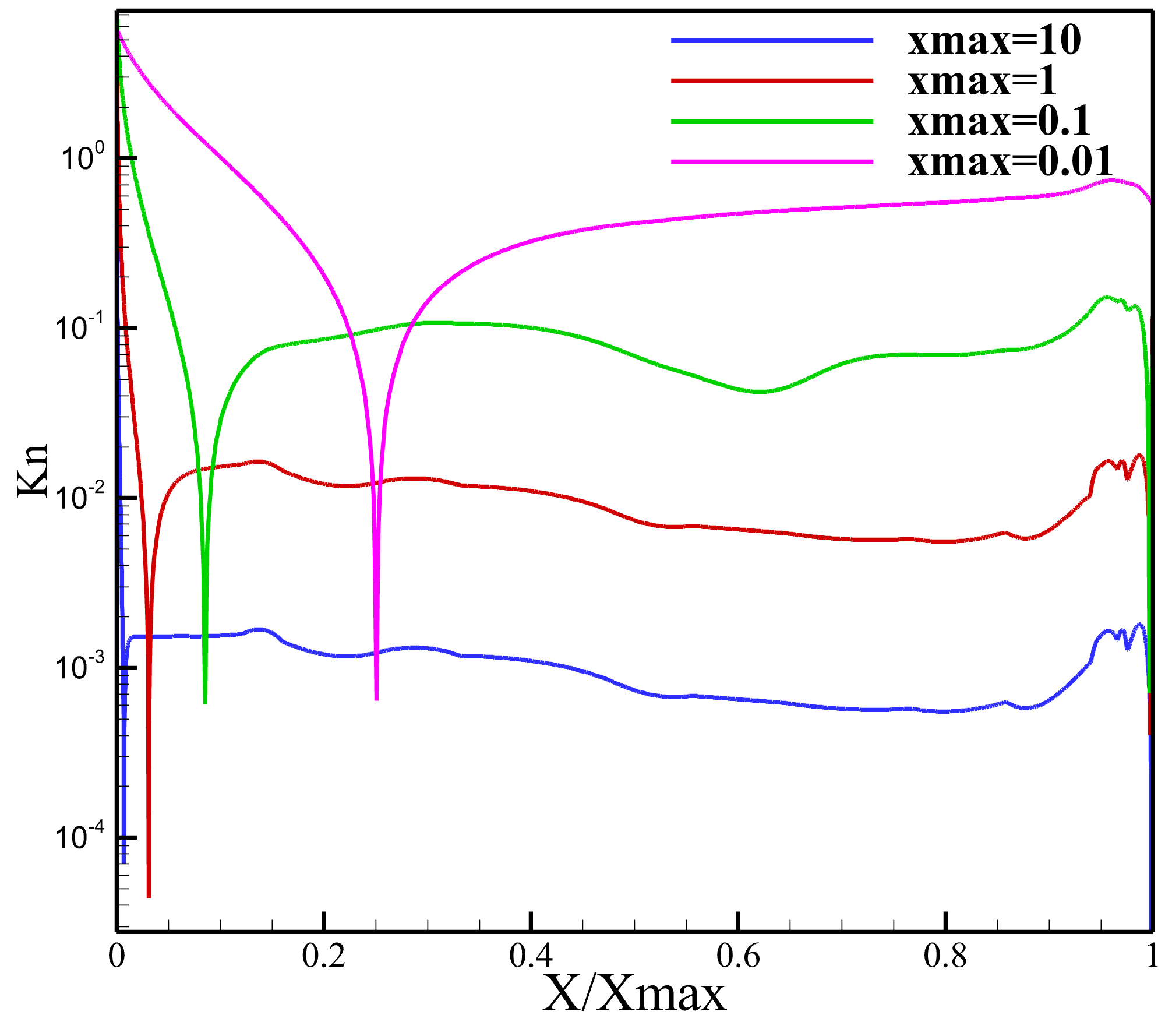}}
\quad
\subfloat[]{\includegraphics[width=0.4\textwidth, clip = true]{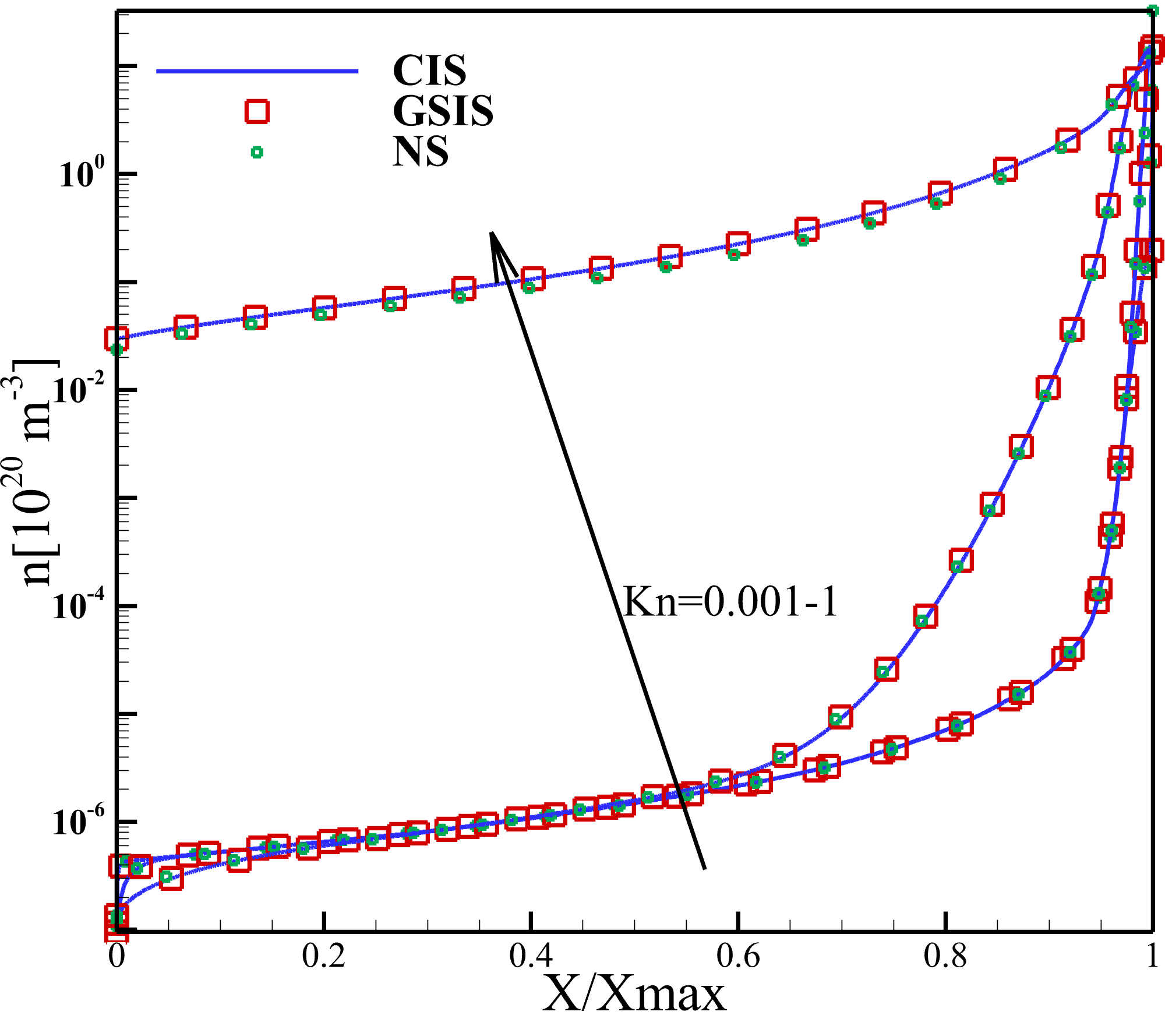}}
\\
\subfloat[]{\includegraphics[width=0.4\textwidth, clip = true]{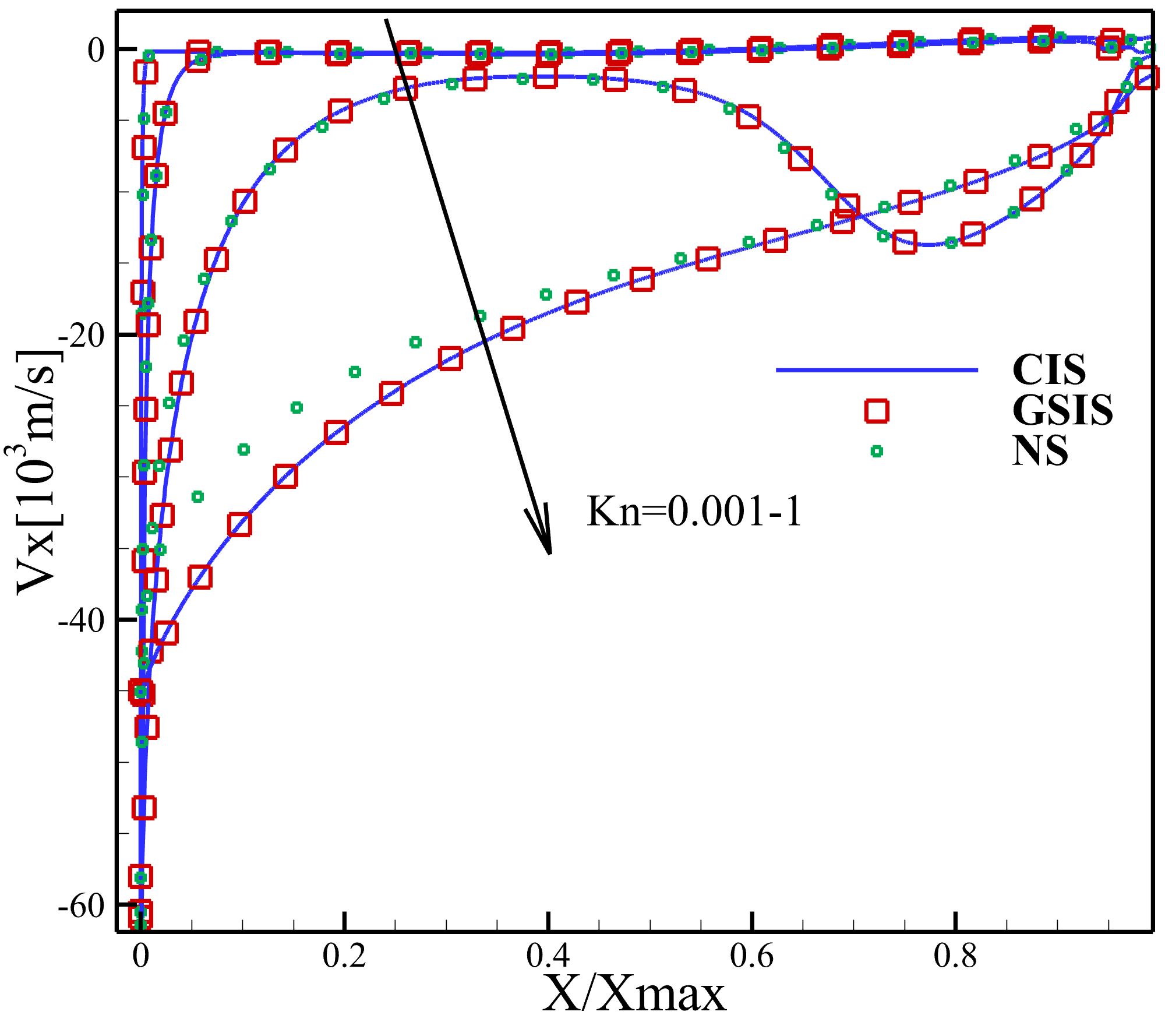}}
\quad
\subfloat[]{\includegraphics[width=0.4\textwidth, clip = true]{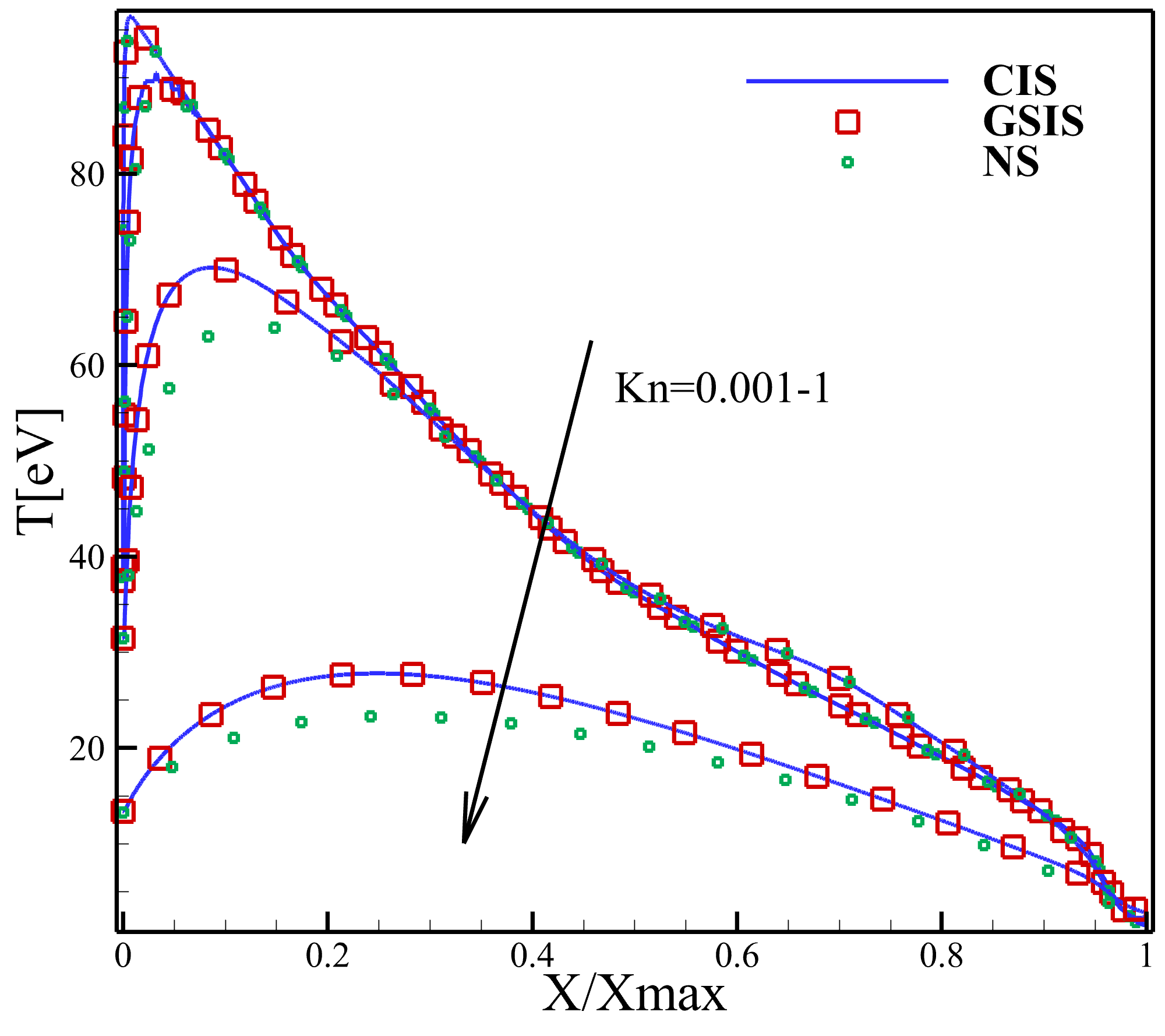}}
\caption{(a) The local Knudsen numbers for different computational domain lengths. (b-d) The density, temperature, and velocity. Along the error, each line corresponds to Kn=0.01, 0.1, 1, and 10. }   
    \label{fig:compare_NS}
\end{figure}

In order to illustrate the shortcomings of the NSF equations in the context of rarefied gas dynamics and to substantiate the precision of GSIS, a series of simulations have been executed across computational domains with lengths of 10 meters, 1 meter, 0.1 meters, and 0.01 meters. Prior to the discussion of the simulation outcomes, we plot the local Knudsen numbers~\eqref{eq:kn}. As depicted in the first subplot of Figure \ref{fig:compare_NS}, the Knudsen numbers vary by nearly five orders of magnitude, underscoring the inherently multi-scale character of the problem. Additionally, it is observed that as the size of the computational domain grows, the average Knudsen number diminishes. In the subsequent sections, the average Knudsen number will be employed to epitomize the findings for each respective domain. Notably, when the maximum domain length $x_{\text{max}}$ is set to 10 meters, the Knudsen number $\text{Kn}$ is 0.001, and when $x_{\text{max}}$ is reduced to 0.01 meters, $\text{Kn}$ escalates to 1.

Figure \ref{fig:compare_NS} illustrates that as the length of the computational domain diminishes, the outcomes derived from the NS equations increasingly diverge from those of the CIS. This divergence is attributable to the breakdown of the continuum hypothesis at elevated Knudsen numbers. This phenomenon is especially pronounced when the Knudsen number $\text{Kn}=1$. Furthermore, Figure \ref{fig:compare_NS} indicates that with the augmentation of the Knudsen number, the distributions of density, velocity, and temperature tend toward greater uniformity. The consistency between the results yielded by GSIS and CIS across various Knudsen numbers serves to validate the precision of the GSIS methodology.

\begin{figure}[t] 
     \centering
\subfloat{{\includegraphics[width=0.45\textwidth, clip = true]{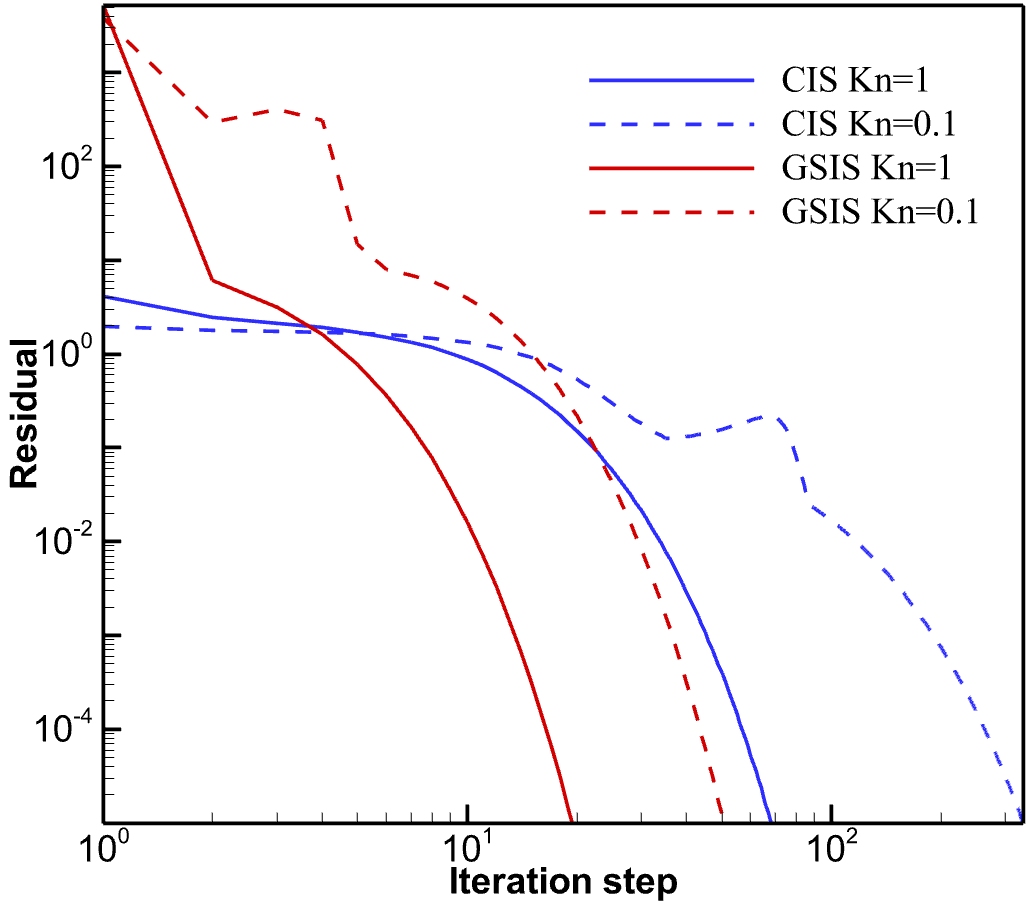}}}\quad
\subfloat{{\includegraphics[width=0.45\textwidth, clip = true]{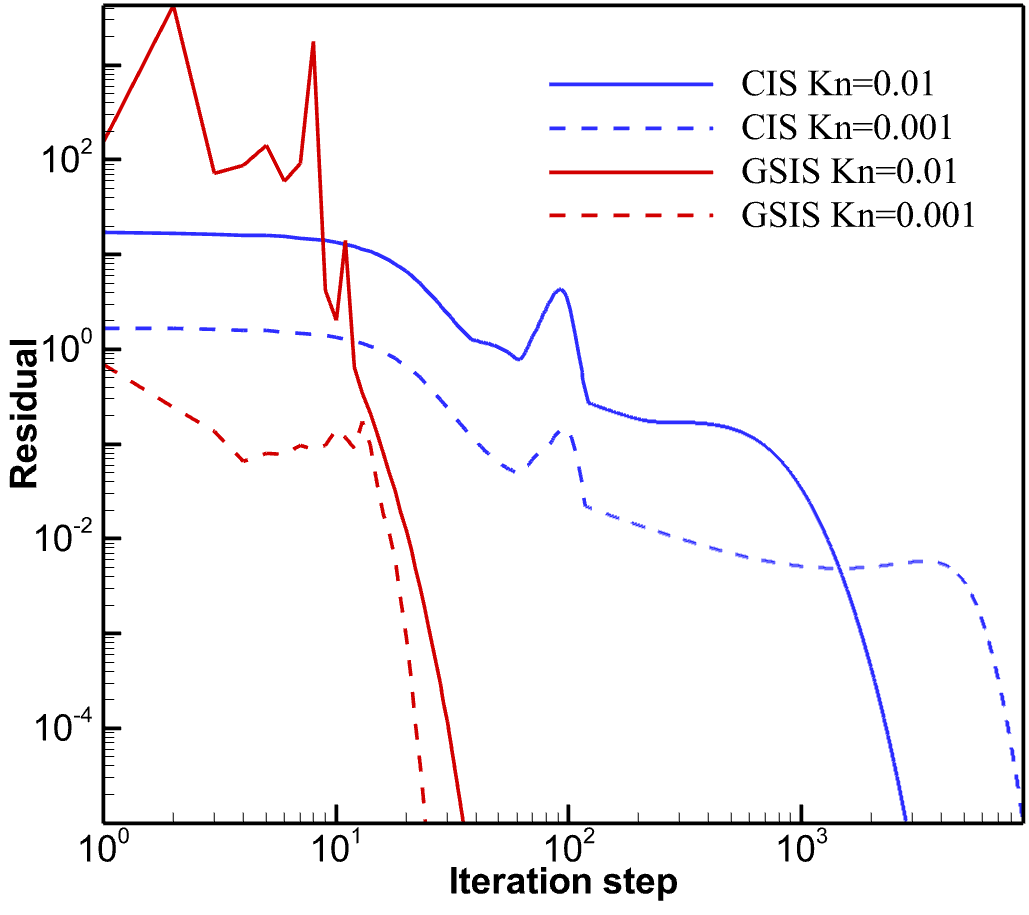}}}
\caption{Error reduction curves of the GSIS and CIS methods at different Knudsen numbers.}   
    \label{fig:err}
\end{figure}

\begin{figure}[t]
    \centering
    \includegraphics[width=0.5\linewidth]{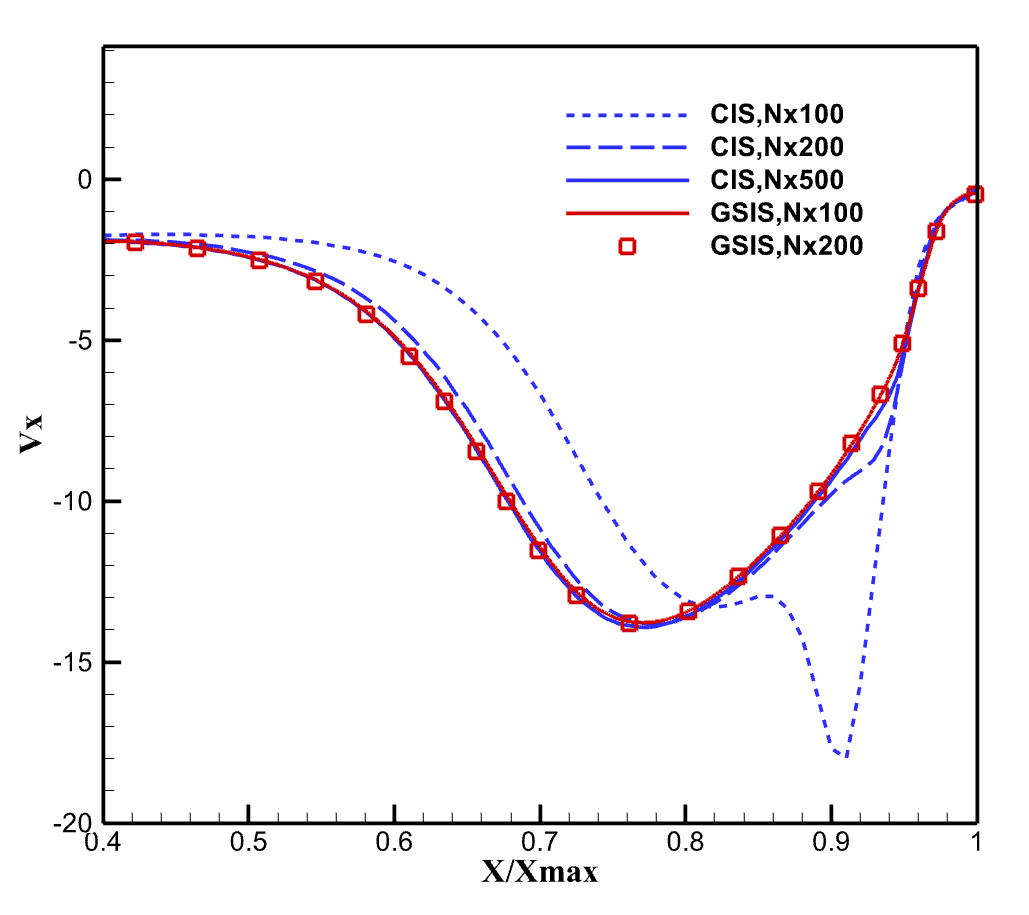}
    \caption{The convergence test in spatial cell numbers.  $N_x$ denotes the number of grid cells in the x-direction, and the grid is uniformly distributed.}
        \label{fig:AP}
\end{figure}

Figure \ref{fig:err} delineates the reduction in error throughout the iterative processes within both the CIS and GSIS methods. At a Knudsen number of $\text{Kn} = 1$, the CIS approach concludes the computation in roughly 70 iterations, whereas the GSIS approach accomplishes the same in just over 20 iterations, realizing a speedup of nearly fourfold. However, as the Knudsen number diminishes, the CIS method experiences a substantial increase in the number of iterations. At $\text{Kn} = 0.1$, the speedup factor is around seven. When the Knudsen number further reduces to $\text{Kn} = 0.01$, CIS demands over 2800 iterations to achieve convergence, whereas GSIS, due to its coupled macroscopic internal iterations that accelerate information exchange within the flow field, converges in just over 30 iterations. With an additional decrease in the Knudsen number to $\text{Kn} = 0.001$, CIS necessitates nearly 9000 iterations to converge, while GSIS still manages to converge in just over 20 iterations, thereby achieving an acceleration of over 400 times in terms of iteration steps.

To assess the asymptotic-preserving characteristic of GSIS, we examine the performance of both the CIS and GSIS methods across various grid resolutions. As depicted in Fig.~\ref{fig:AP}, the CIS method exhibits a significant deviation from the accurate solution when the grid resolution is set to 100 points. With an increased resolution of 200 grid points, there is a marked improvement in the CIS results, yet a discernible deviation from the correct solution persists. On the other hand, GSIS delivers results that are nearly indistinguishable from those obtained by the CIS method with 500 grid points, even when the grid resolution is as low as 100 points. This comparison underscores the robustness and efficiency of GSIS in maintaining solution accuracy across different grid resolutions.

\begin{figure}[!ht]
 \centering
\subfloat{{\includegraphics[width=0.33\textwidth,  clip = true]{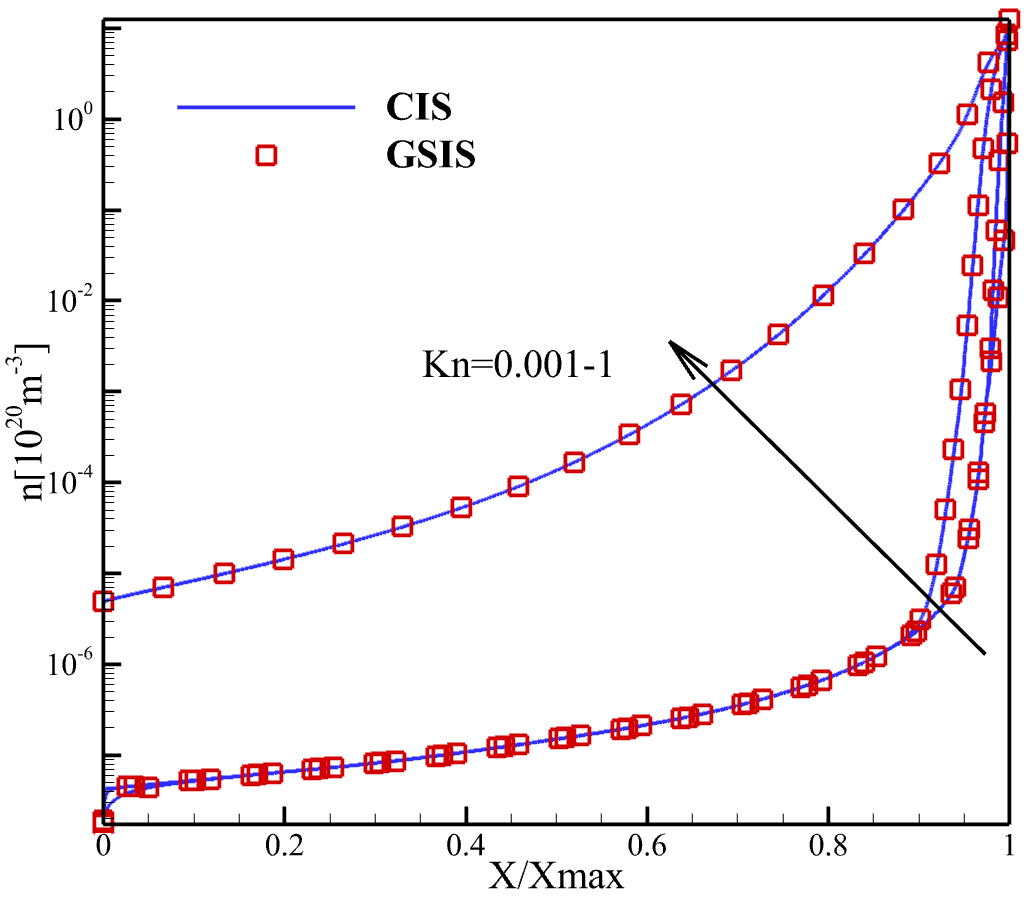}}}
\subfloat{{\includegraphics[width=0.33\textwidth,  clip = true]{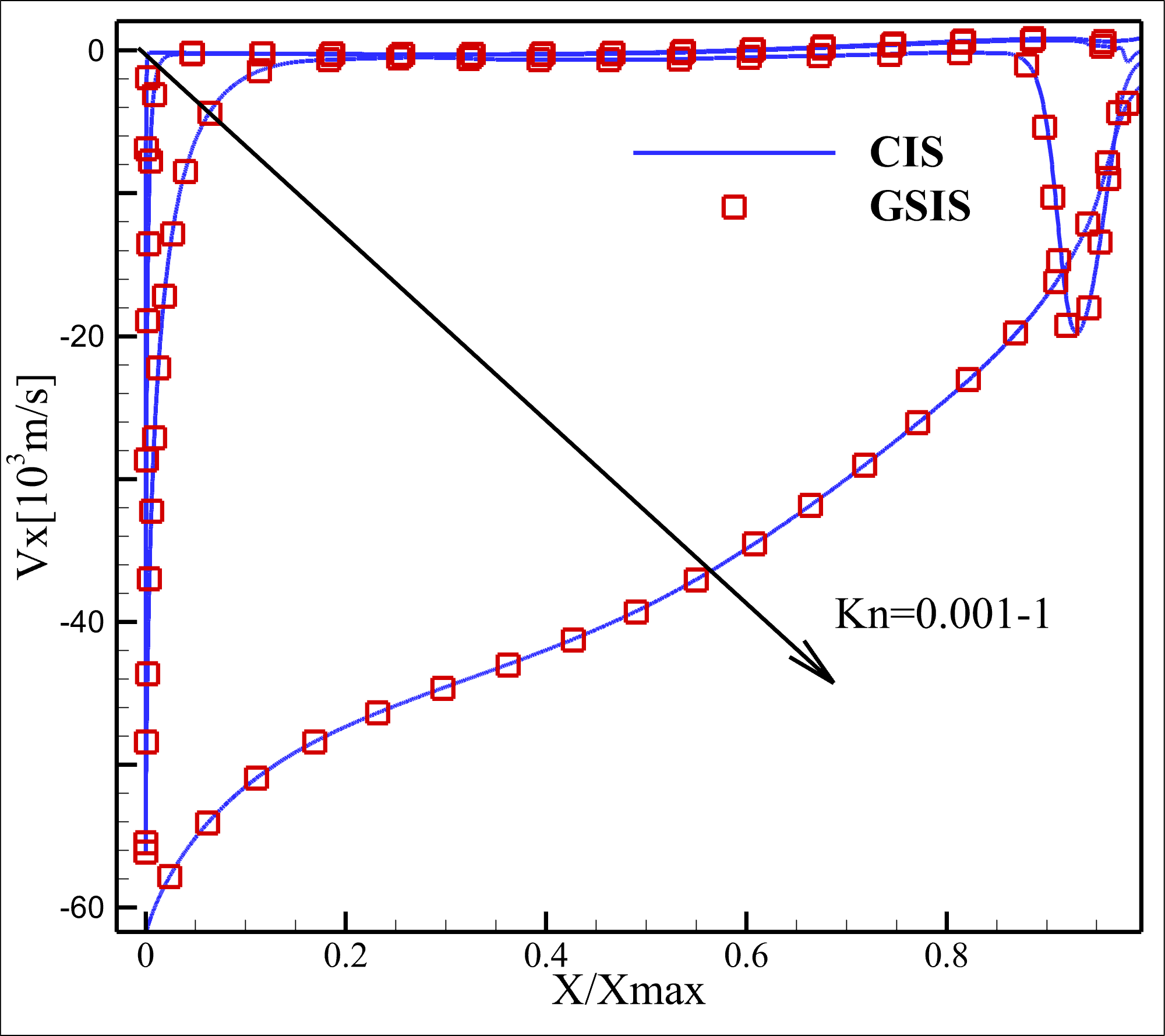}}}
\subfloat{{\includegraphics[width=0.33\textwidth,  clip = true]{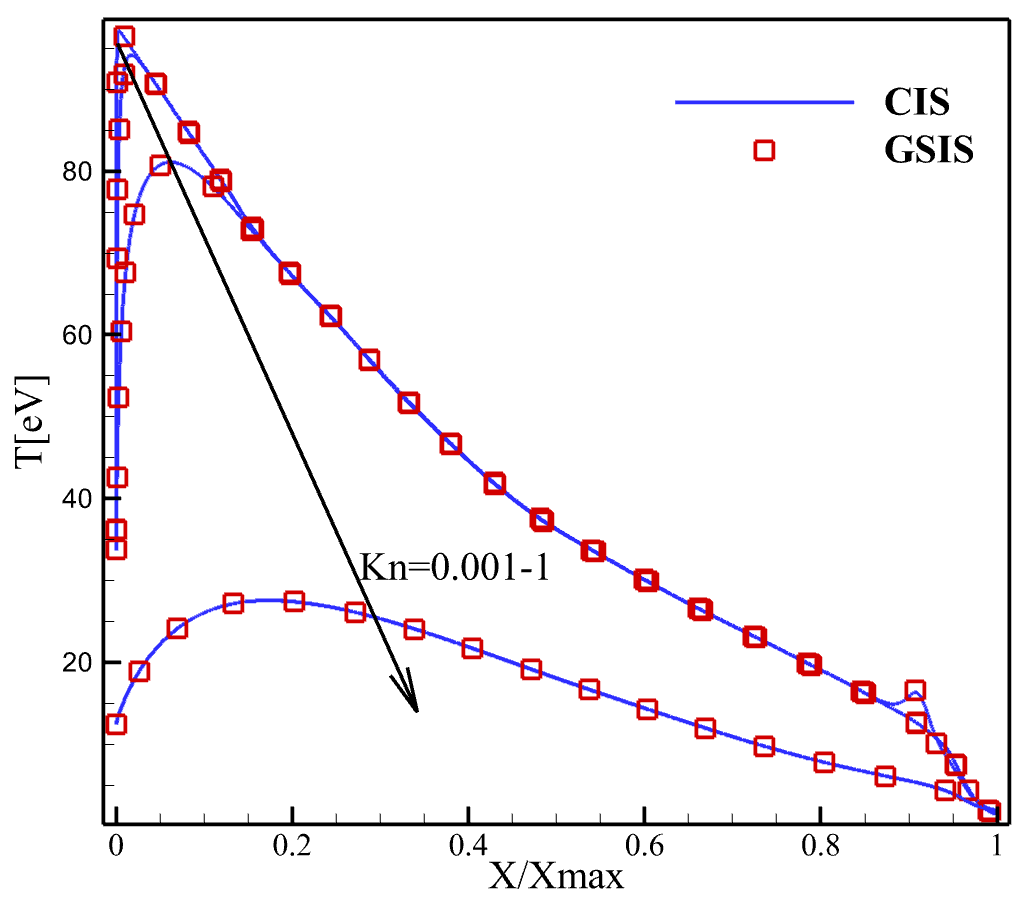}}}
\\
\subfloat{{\includegraphics[width=0.33\textwidth,  clip = true]{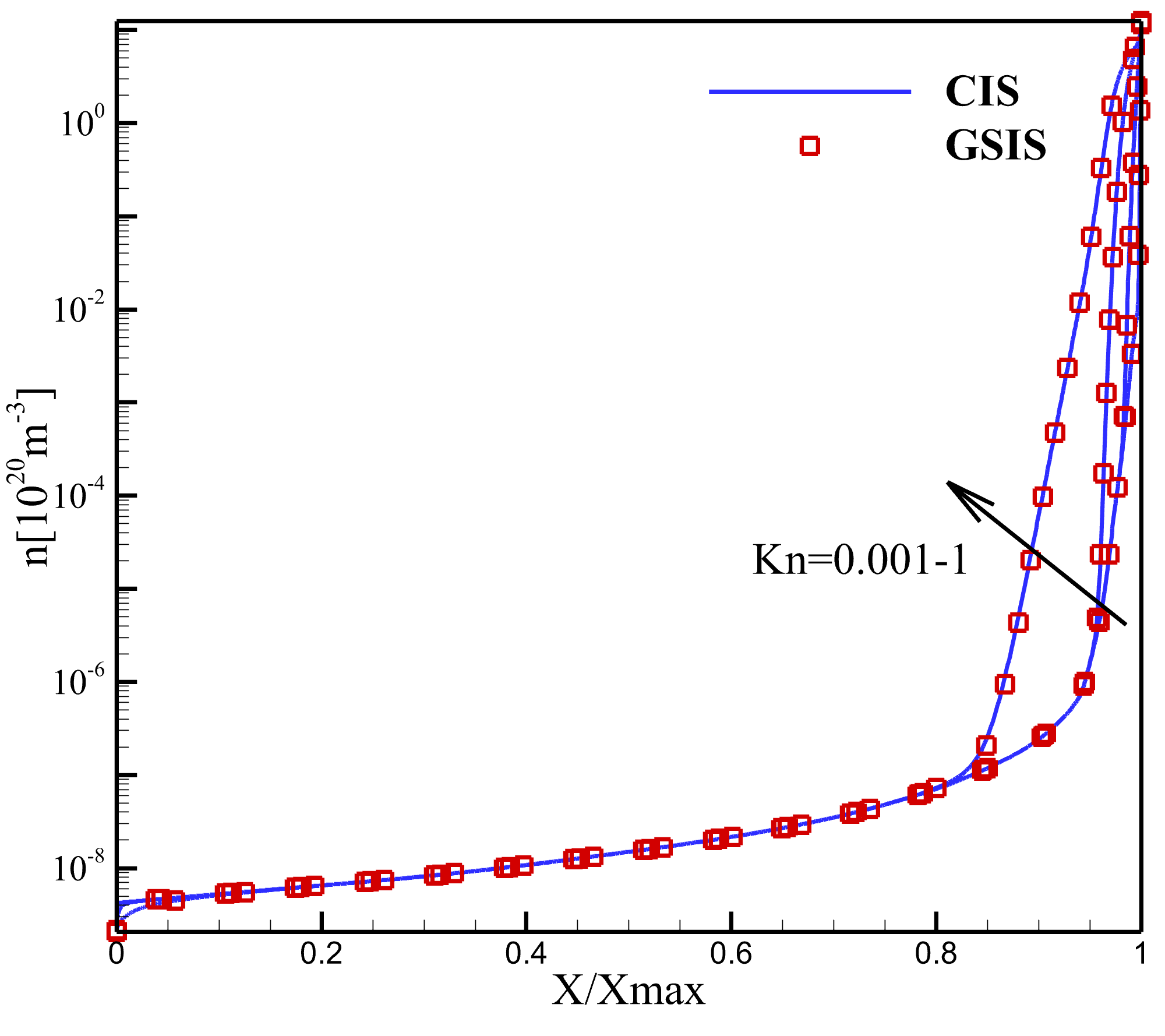}}}
\subfloat{{\includegraphics[width=0.33\textwidth,  clip = true]{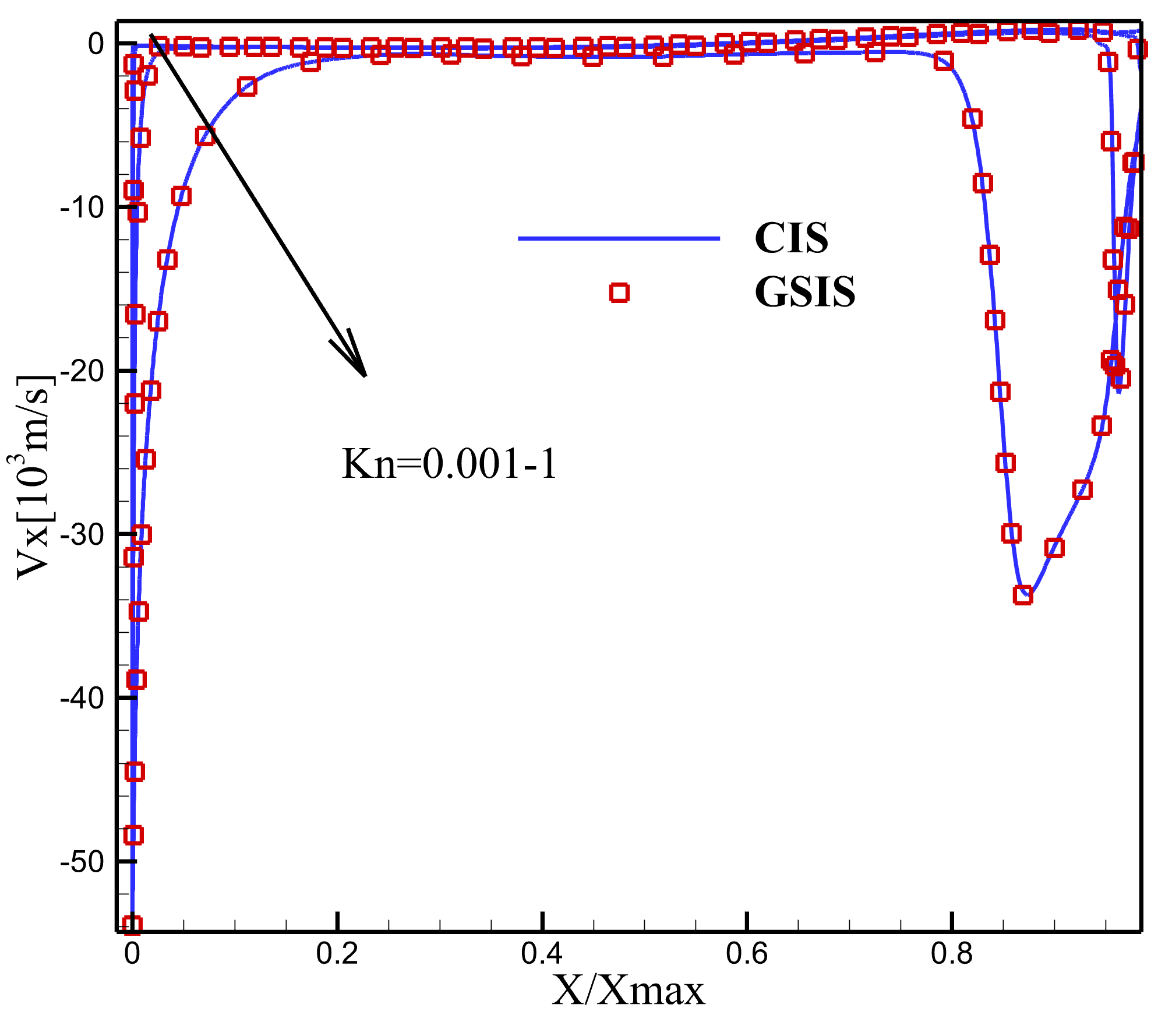}}}
\subfloat{{\includegraphics[width=0.33\textwidth,  clip = true]{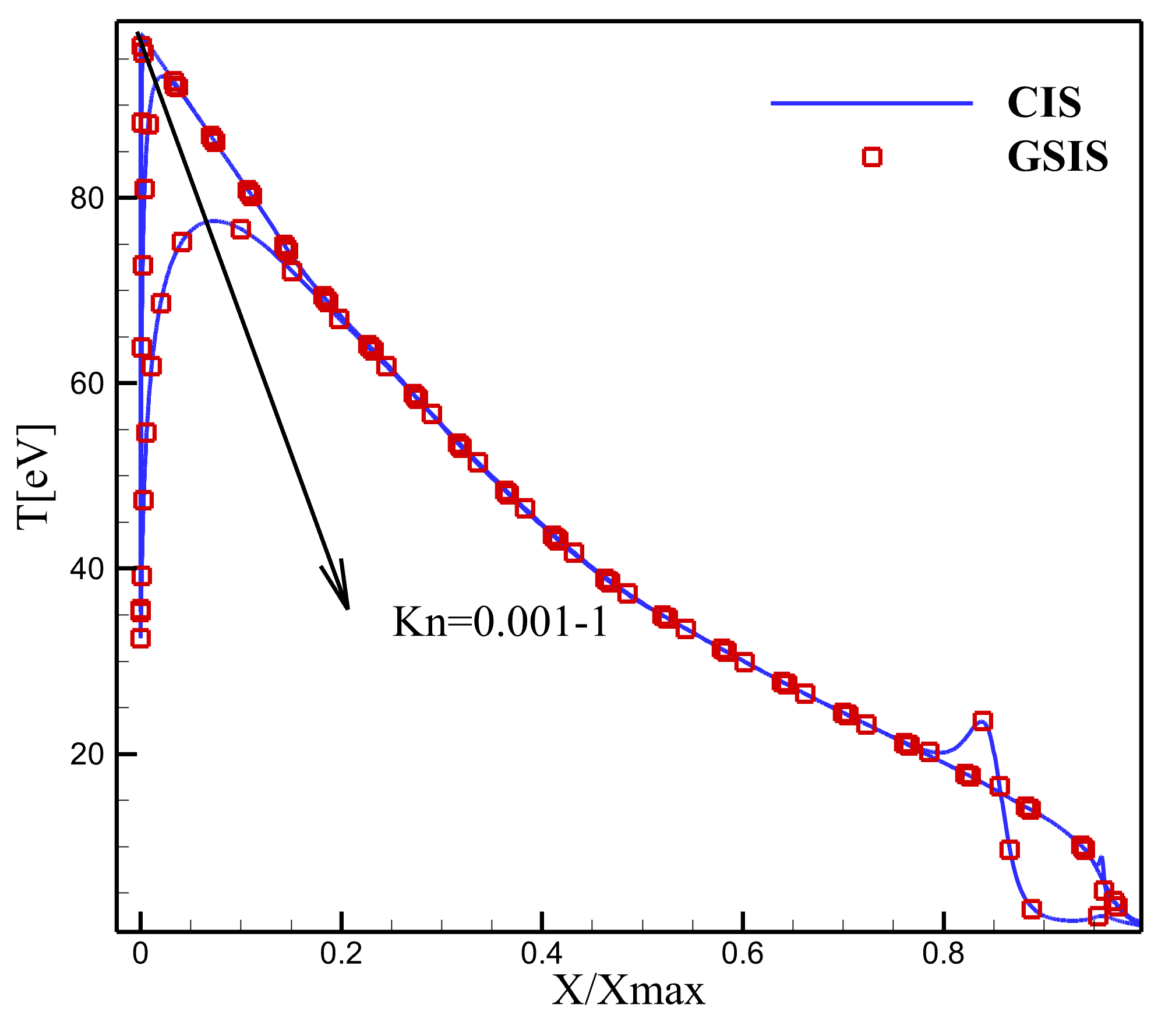}}}
  \caption{The density, velocity, and temperature distributions at different Knudsen numbers after a (first row) 10-fold and (second row) 100-fold increase in $K_D$. }
  \label{fig:10-100KD}
\end{figure}

Finally, we increase the ionization coefficient $K_D$ by 10 and 100 times to simulate conditions where the ionization process becomes dominant. The final converged solutions are shown in Fig.~\ref{fig:10-100KD}. A comparison with Fig.~\ref{fig:compare_NS} clearly reveals a significant decrease in particle density. This is because the increasing the ionization coefficient greatly enhances the proportion of the ionization process, leading to a substantial conversion of neutral particles into ions and, consequently, a noticeable decrease in density. Additionally, Fig.~\ref{fig:10-100KD} demonstrates that after increasing $K_D$ by 10 and 100 times, the GSIS method still achieves results that are nearly identical to those obtained by the CIS method across various Knudsen numbers. The final acceleration performance, shown in Table~\ref{tab:10-100KD}, indicates that there is a consistent acceleration effect across both high and low Knudsen numbers, with the acceleration effect becoming more pronounced as the Knudsen number decreases. These results confirm the efficiency and versatility of the GSIS algorithm in handling problems with varying degrees of rarefaction and dominant processes.

\begin{table}[t]
    \centering
    \caption{The number of iterations required for convergence between the CIS and GSIS when $K_D$ in Eq.~\eqref{eq:K} is increased by 10-fold and 100-fold, respectively.} 
    \begin{tabular}{ccccc}
        \hline 
        \multirow{2}{*}{\text{Kn}} & \multicolumn{2}{c}{10 $K_D$} & \multicolumn{2}{c}{100 $K_D$} \\
        \cline{2-5} 
        & CIS & GSIS & CIS & GSIS \\
        \hline 
        0.001 & 4051 & 22 & 952 & 23 \\
        0.01 & 994 & 21 & 404 & 15 \\
        0.1 & 158 & 12 & 91 & 13 \\
        1 & 27 & 14 & 14 & 11 \\
        \hline
    \end{tabular}
    \label{tab:10-100KD}
\end{table}

\begin{figure}[t]
 \centering
\subfloat{{\includegraphics[width=0.47\textwidth,  clip = true]{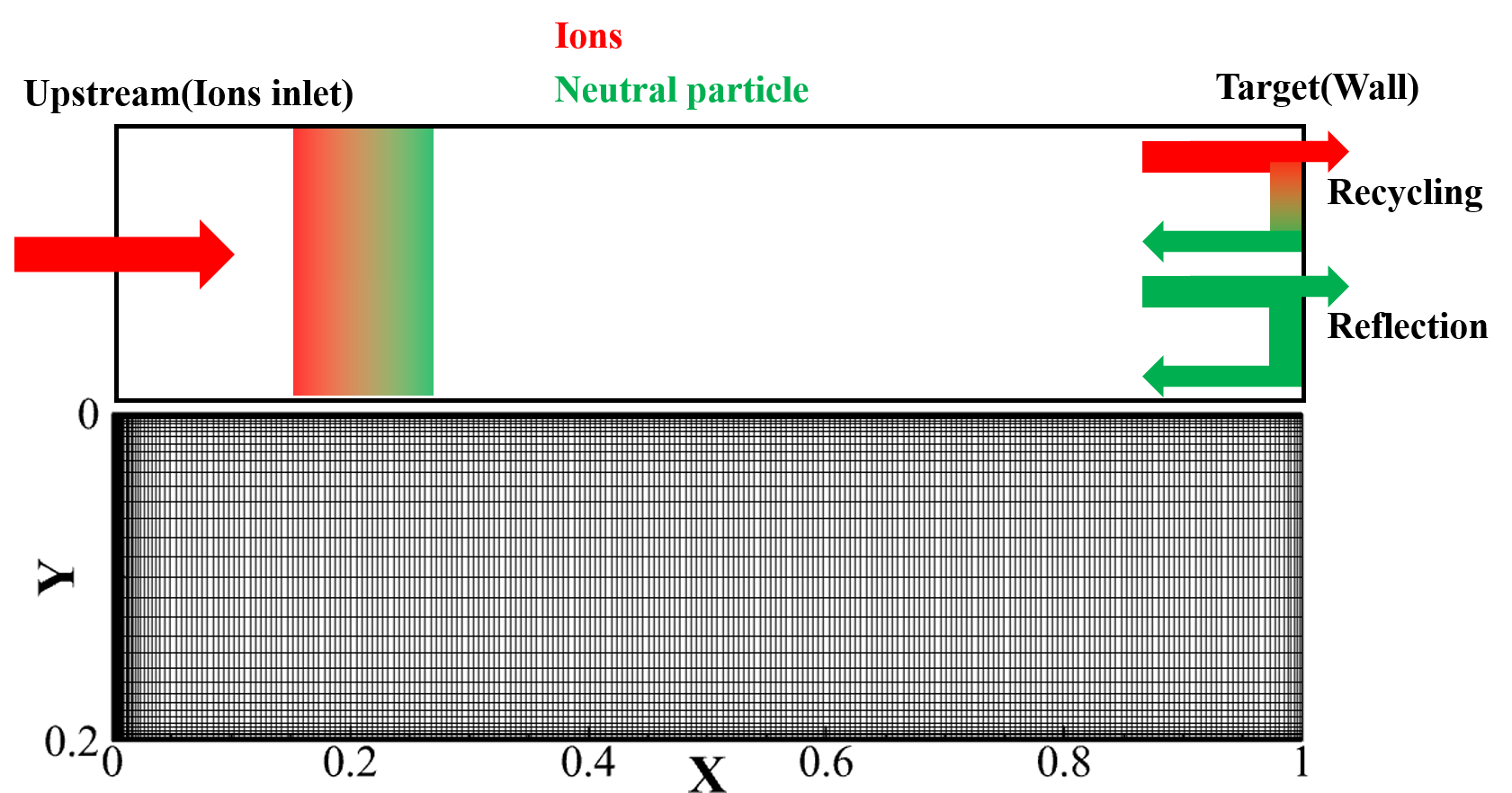}}}
\subfloat{{\includegraphics[width=0.47\textwidth,  clip = true]{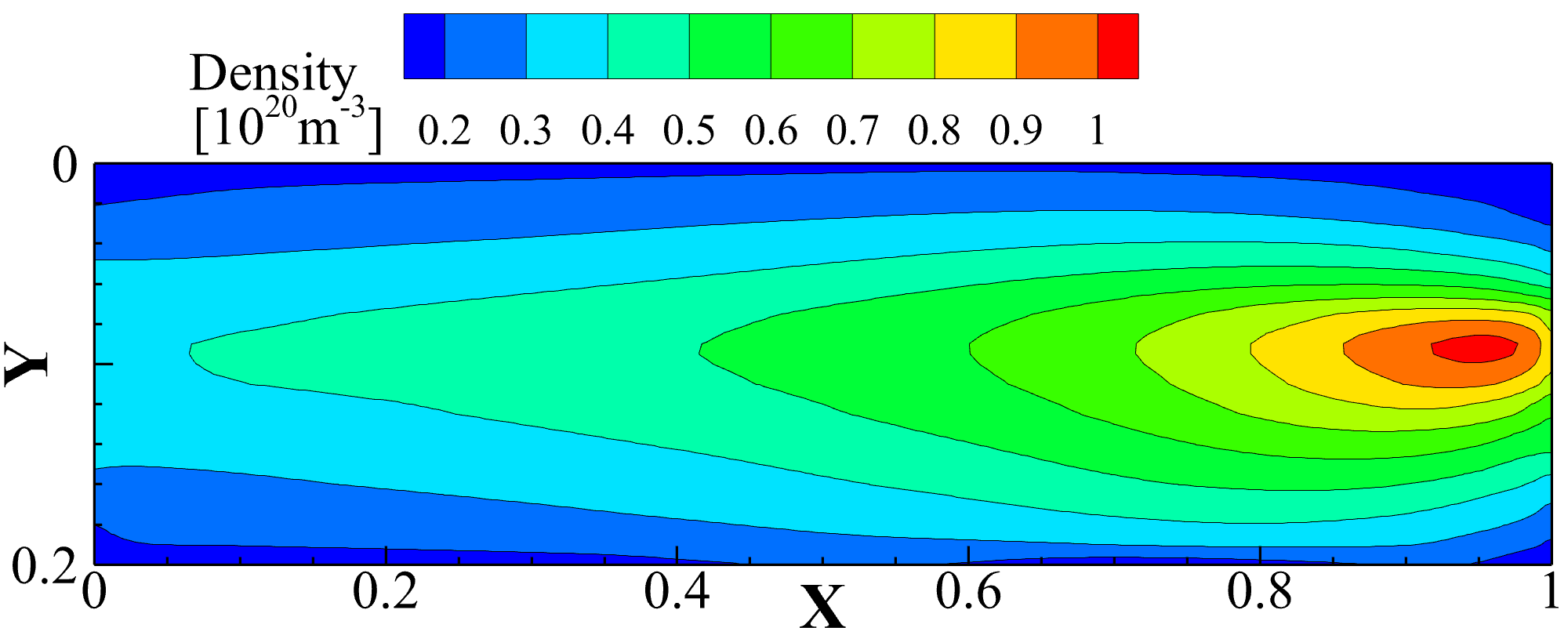}}}
\\
\subfloat{{\includegraphics[width=0.47\textwidth,  clip = true]{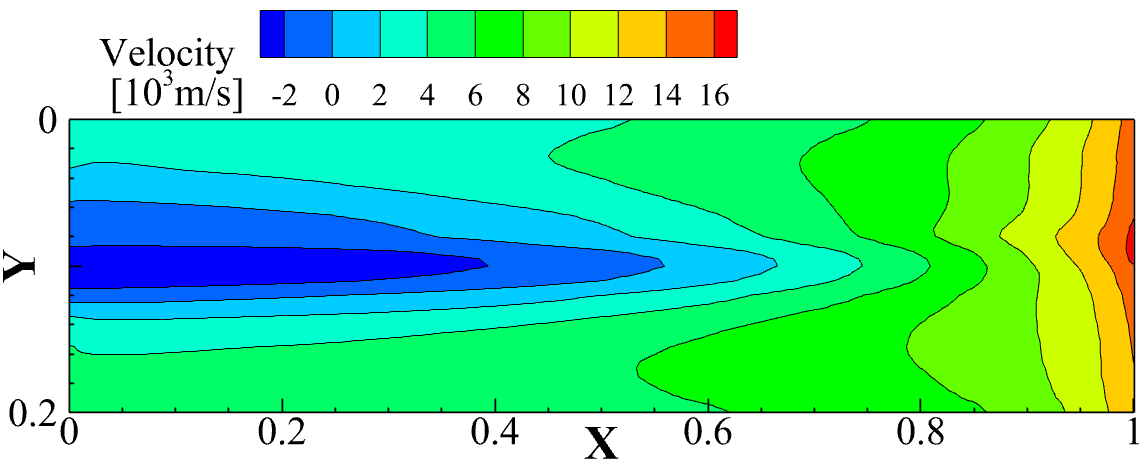}}}
\subfloat{{\includegraphics[width=0.47\textwidth,  clip = true]{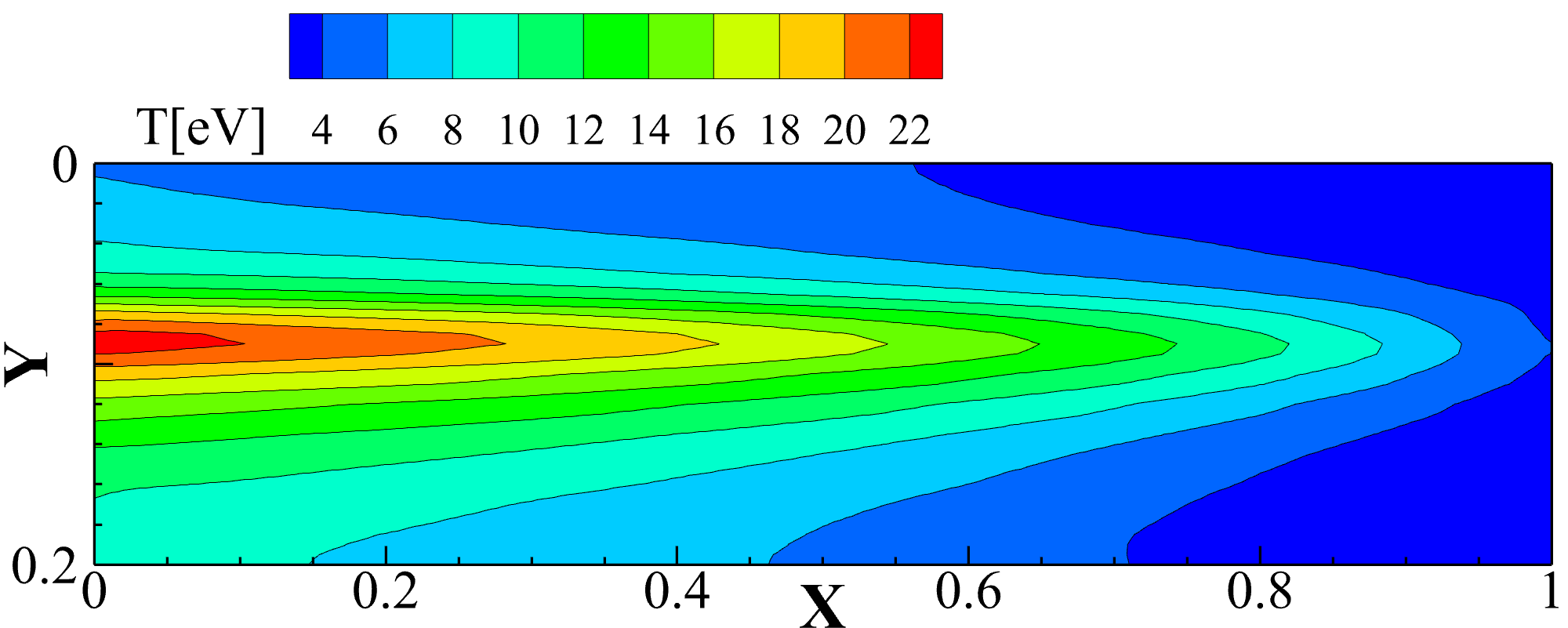}}}
  \caption{The 2D model and mesh, and the background ion density, velocity, and temperature \cite{HorstenJCP2020}.}
  \label{fig:BG_2D}
\end{figure}

\subsection{Two-dimensional edge plasma flow}

Consider the edge plasma problem in two dimensions, where the Knudsen number varies between 1 and 0.001. As shown in Fig.~\ref{fig:BG_2D}, the boundary conditions at the upstream and downstream are similar to those in the one-dimensional case. At \(y = 0.2\), the diffuse reflection boundary condition is implemented, while at \(y = 0\), the diffuse reflection boundary condition with an absorption rate is applied. 
The three bottom panels of Fig.~\ref{fig:BG_2D} illustrate the fixed background plasma state, with the ion density, velocity, and temperature displayed from top to bottom. This background is derived from Ref.~\cite{HorstenJCP2020}, and obtained through interpolation.
The physical domain is discretized using a non-uniform Cartesian grid, with grid spacing being refined in proximity to the walls to capture the boundary layer effects more accurately. A total of $250 \times 40$ grid points are used at each Knudsen number. The velocity space is divided into $40 \times 40$ velocity grid cells. The maximum velocity in the velocity space is given by \(10\sqrt{k_B T_{D,\text{max}}/m}\), where \(T_{D,\text{max}}\) is the maximum ion temperature.

\begin{figure}[p]
     \centering
\subfloat{{\includegraphics[width=0.45\textwidth, clip = true]{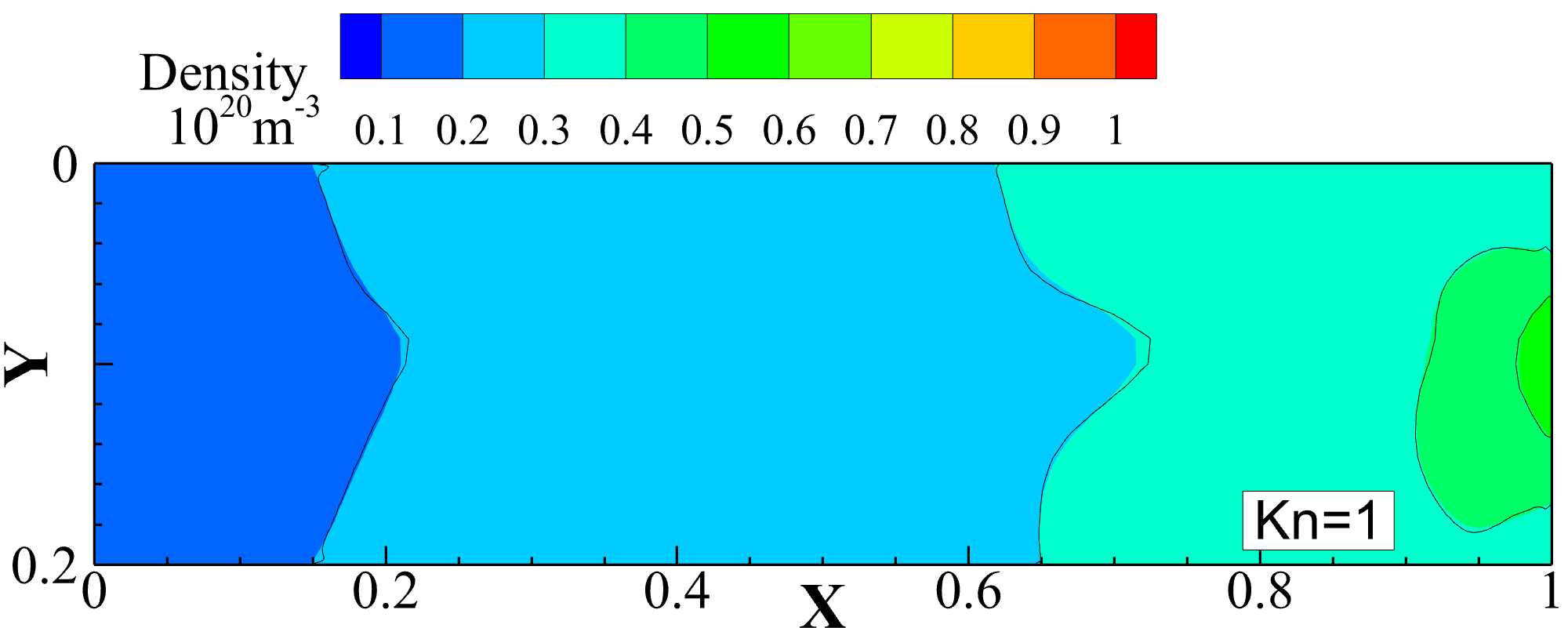}}}
     \subfloat{
{\includegraphics[width=0.45\textwidth,  clip = true]{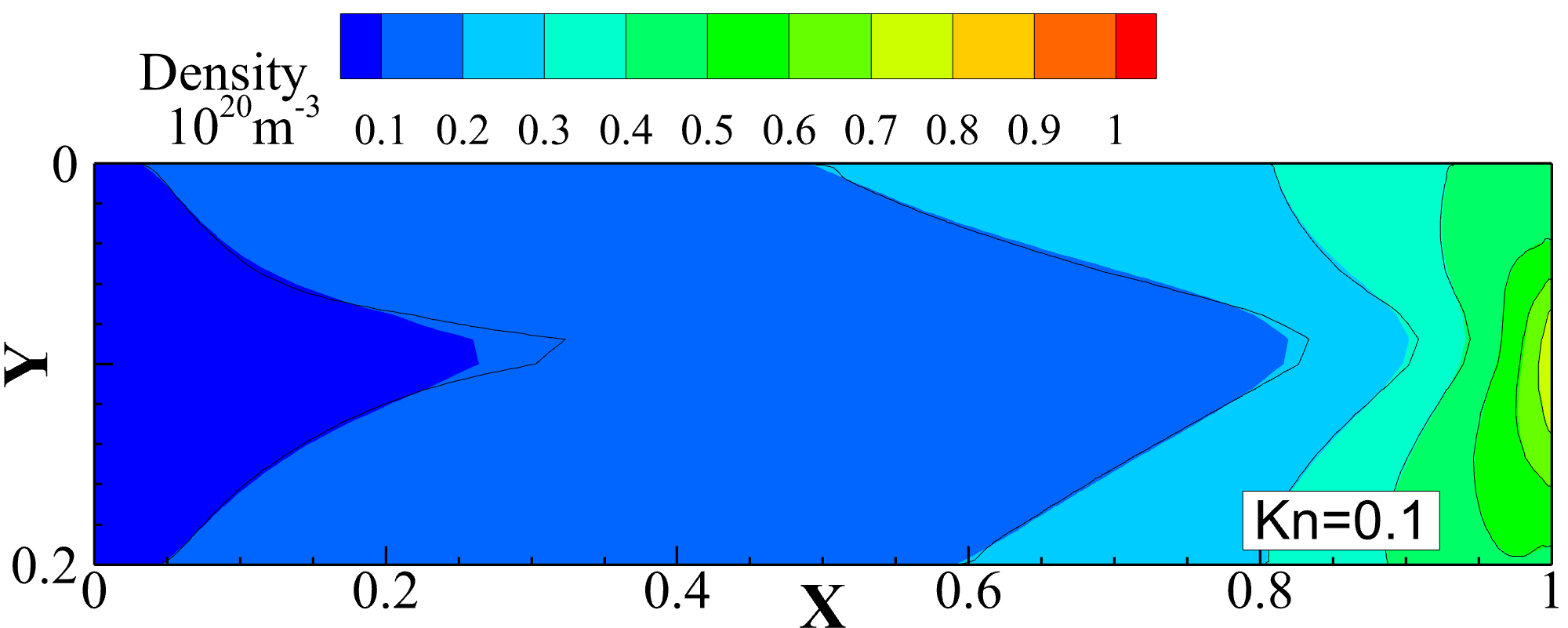}}}
     \\
     \subfloat{
{\includegraphics[width=0.45\textwidth,  clip = true]{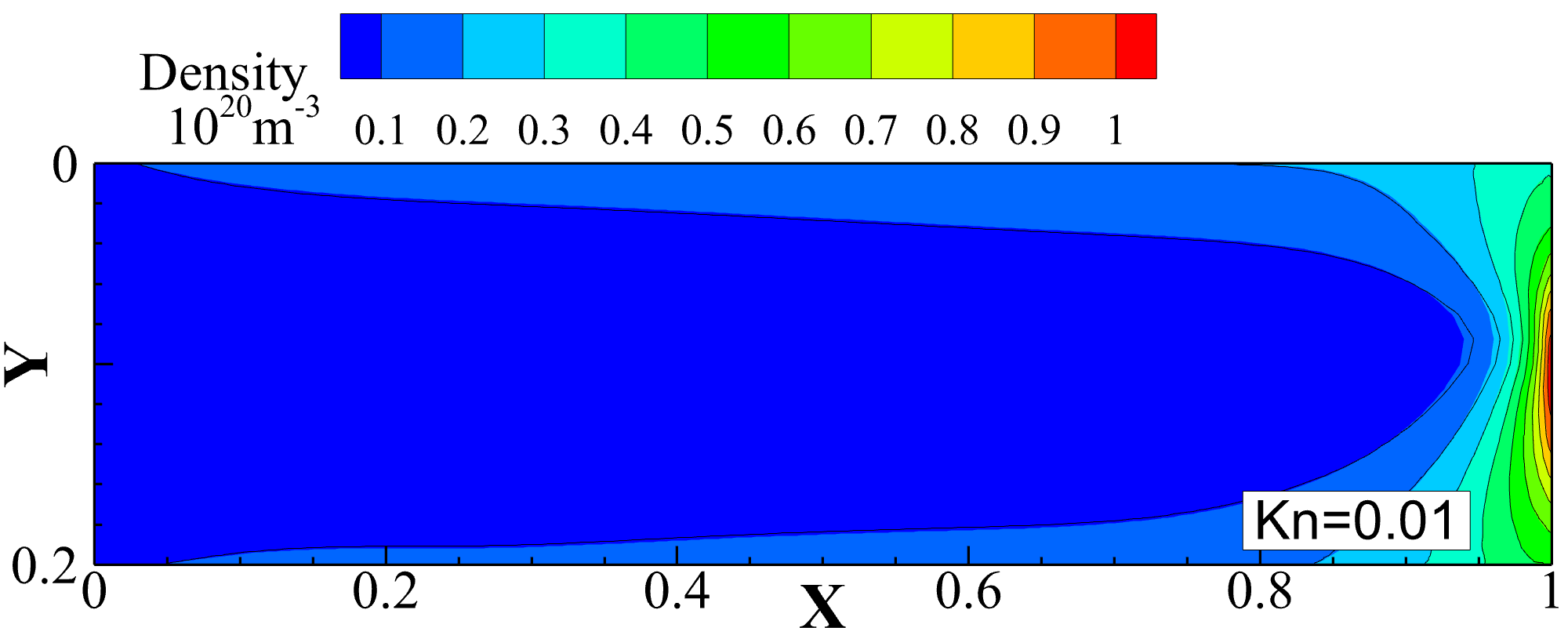}}}
     \subfloat{
{\includegraphics[width=0.45\textwidth,  clip = true]{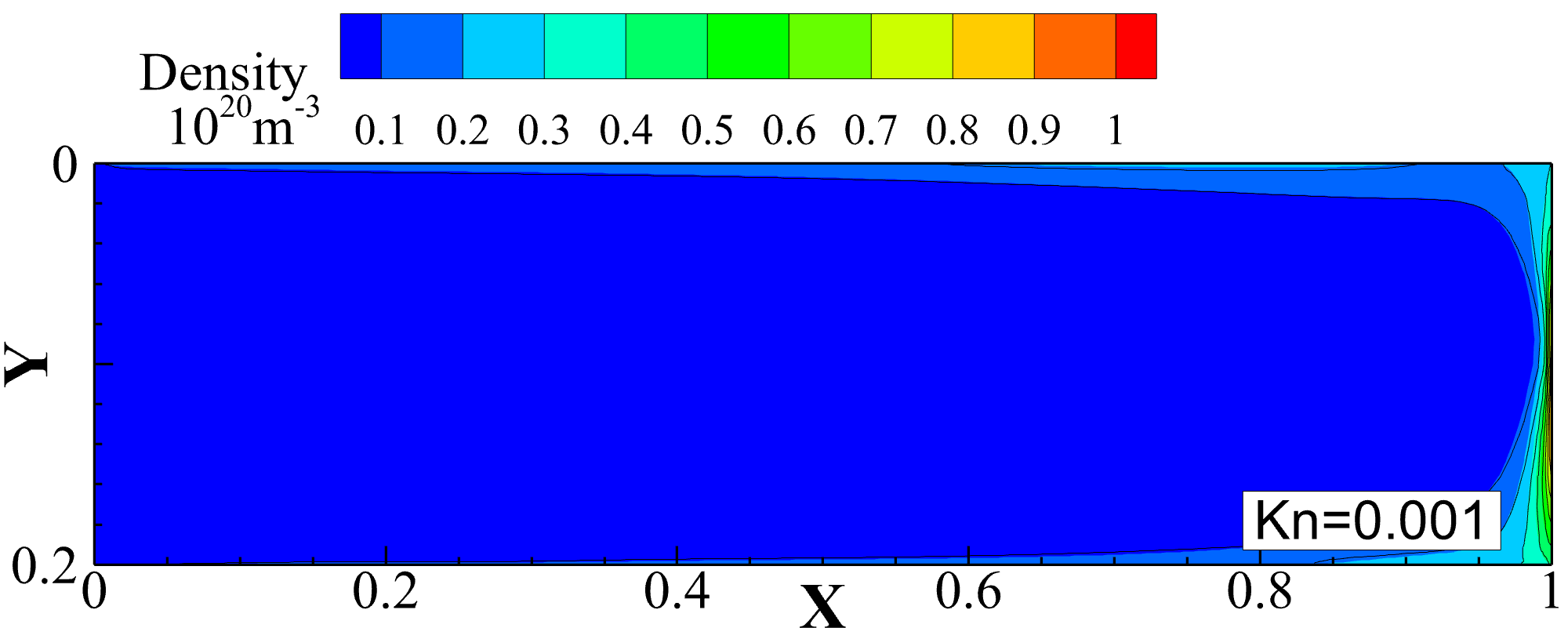}}}
     \\
\subfloat{
{\includegraphics[width=0.45\textwidth, clip = true]{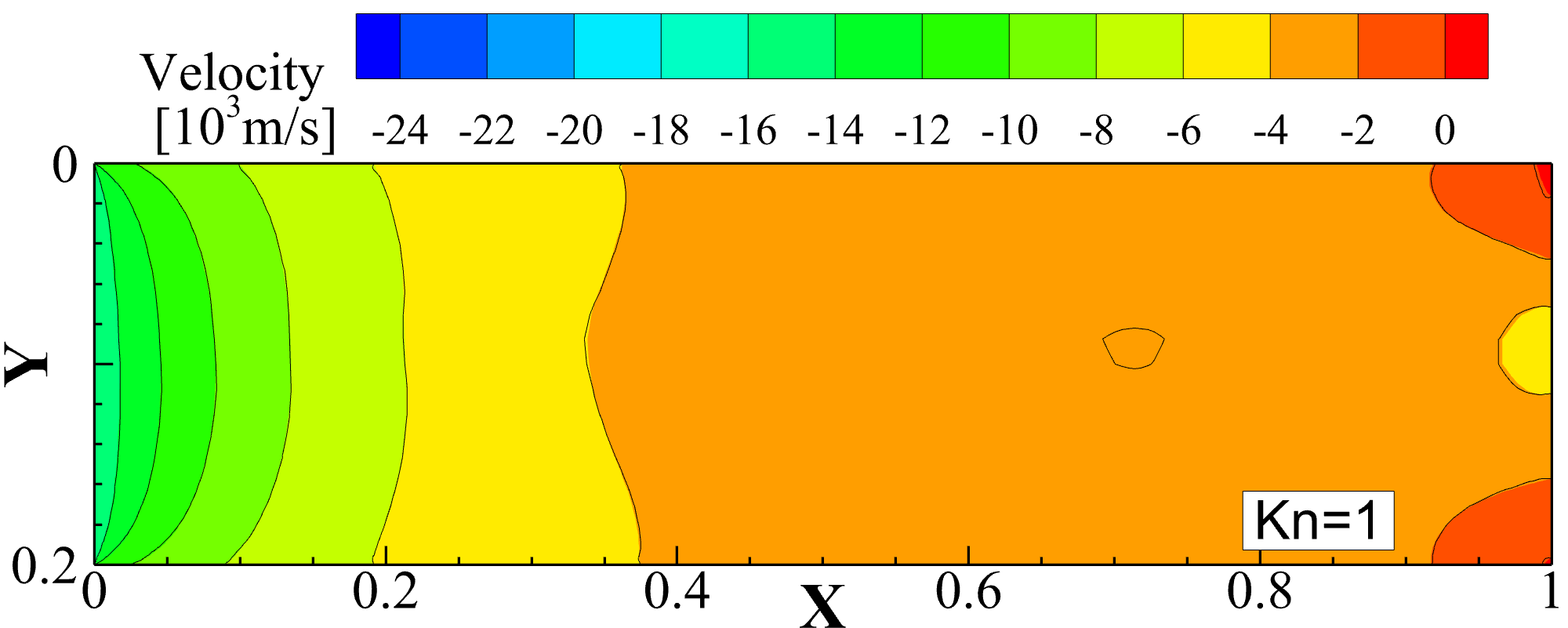}}}
     \subfloat{
{\includegraphics[width=0.45\textwidth,  clip = true]{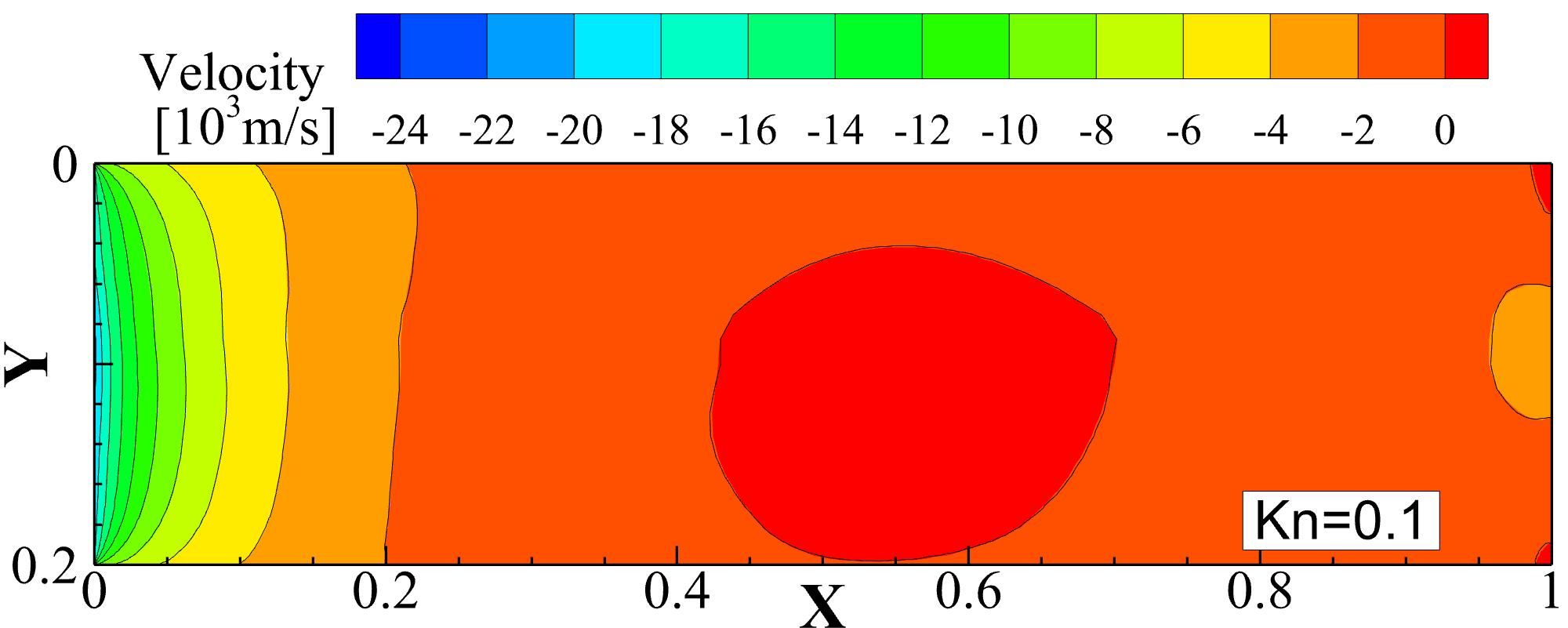}}}
     \\
     \subfloat{
{\includegraphics[width=0.45\textwidth,  clip = true]{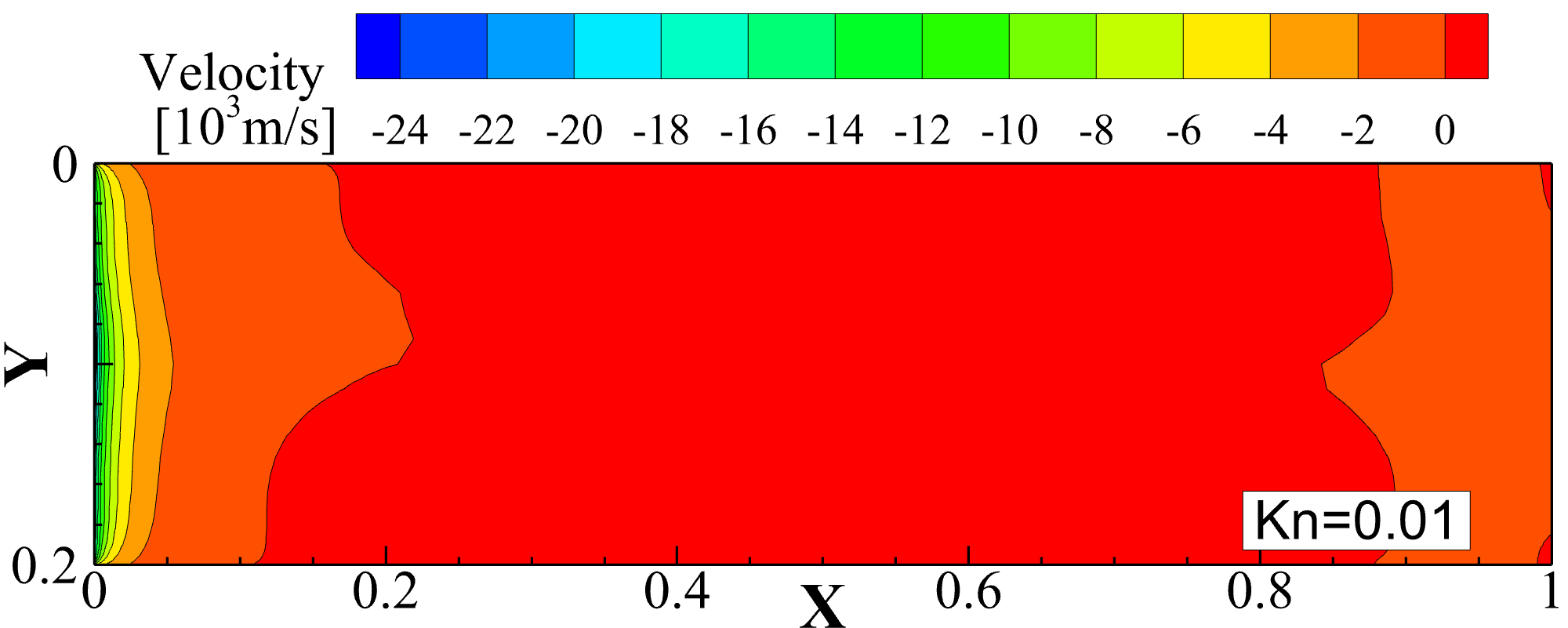}}}
     \subfloat{
{\includegraphics[width=0.45\textwidth,  clip = true]{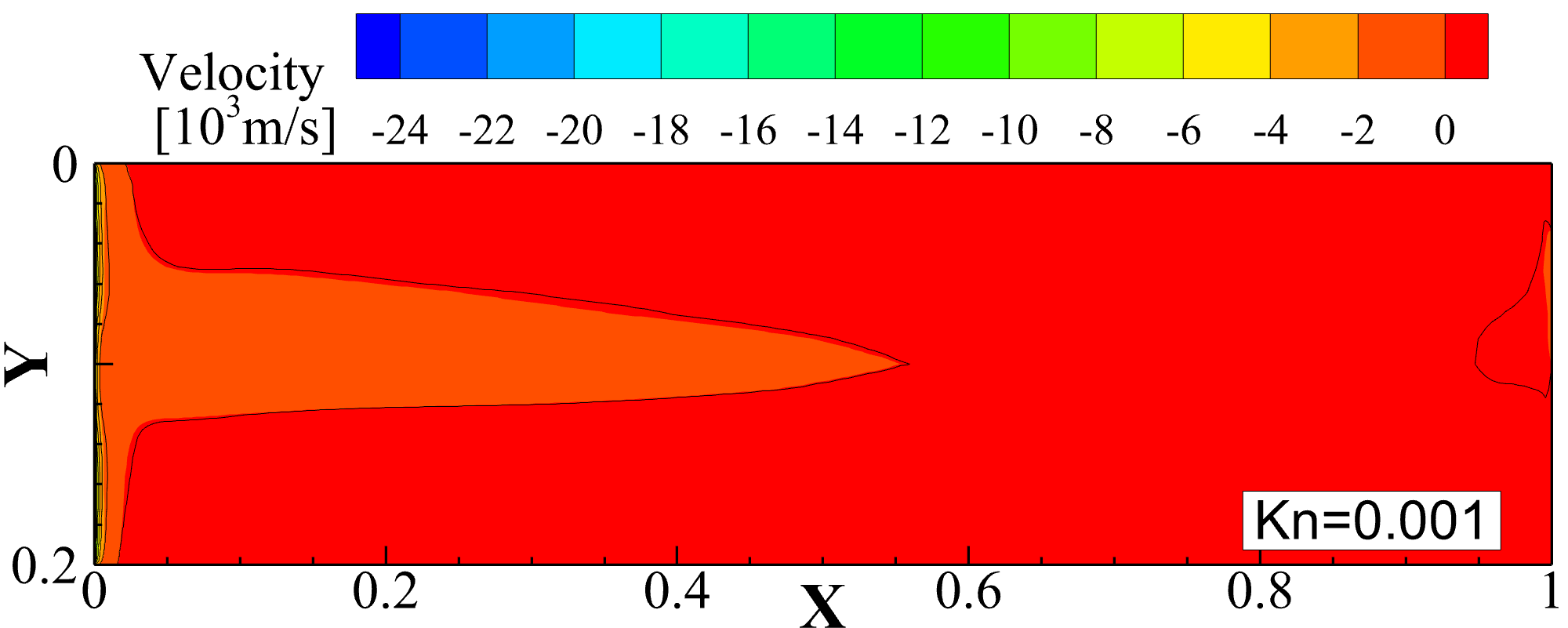}}}
    \\
    \subfloat{
{\includegraphics[width=0.45\textwidth, clip = true]{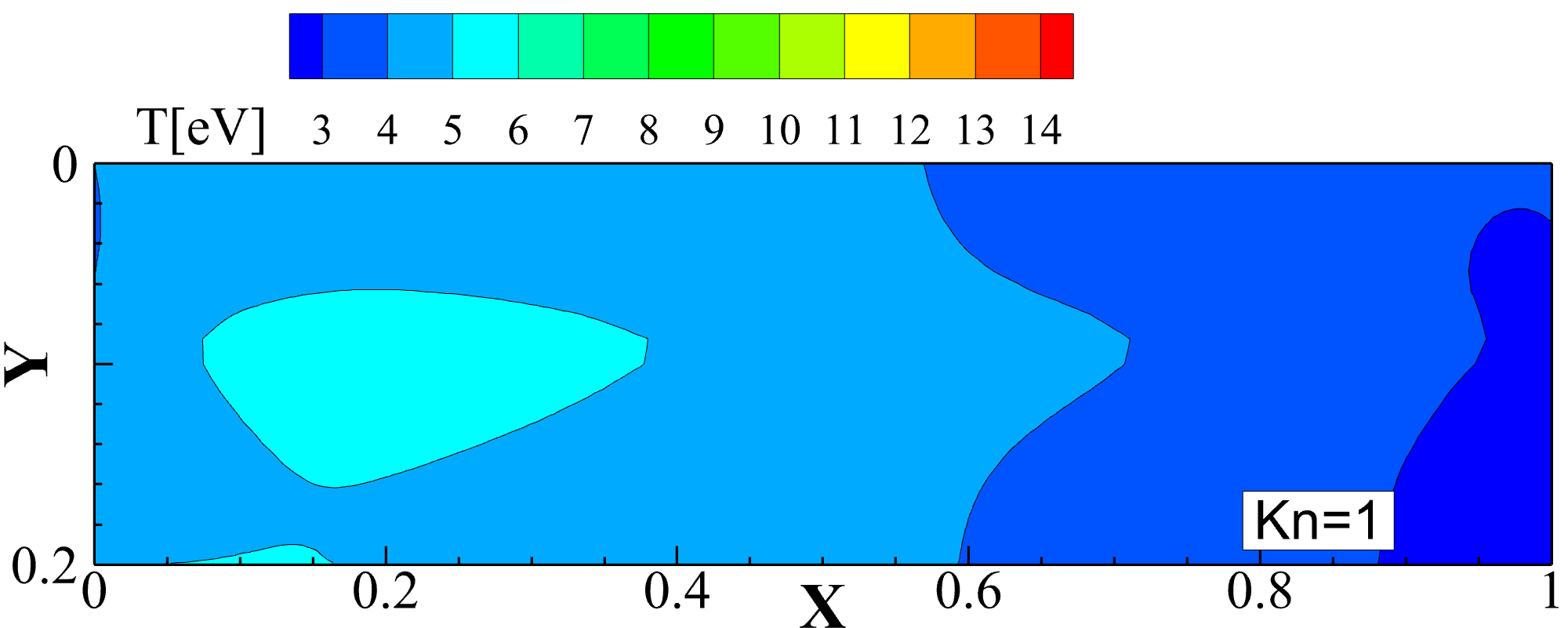}}}
     \subfloat{
{\includegraphics[width=0.45\textwidth,  clip = true]{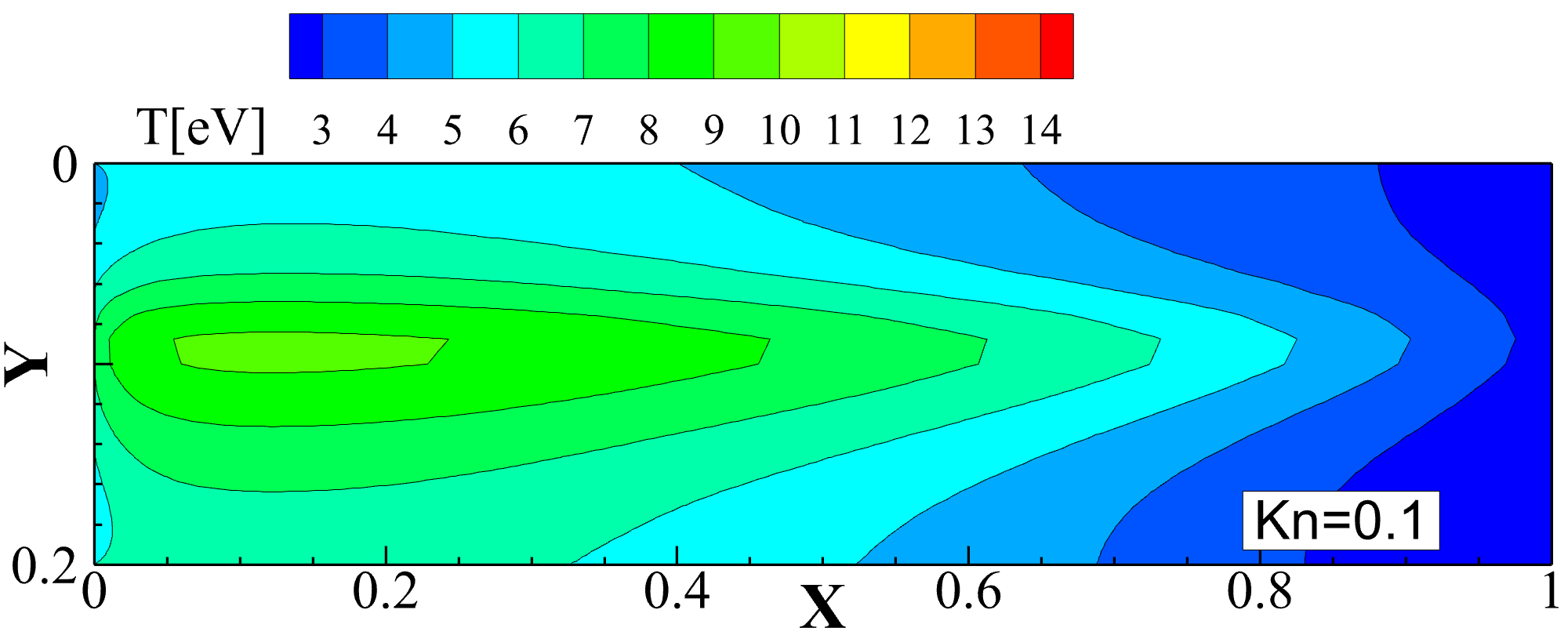}}}
     \\
     \subfloat{
{\includegraphics[width=0.45\textwidth,  clip = true]{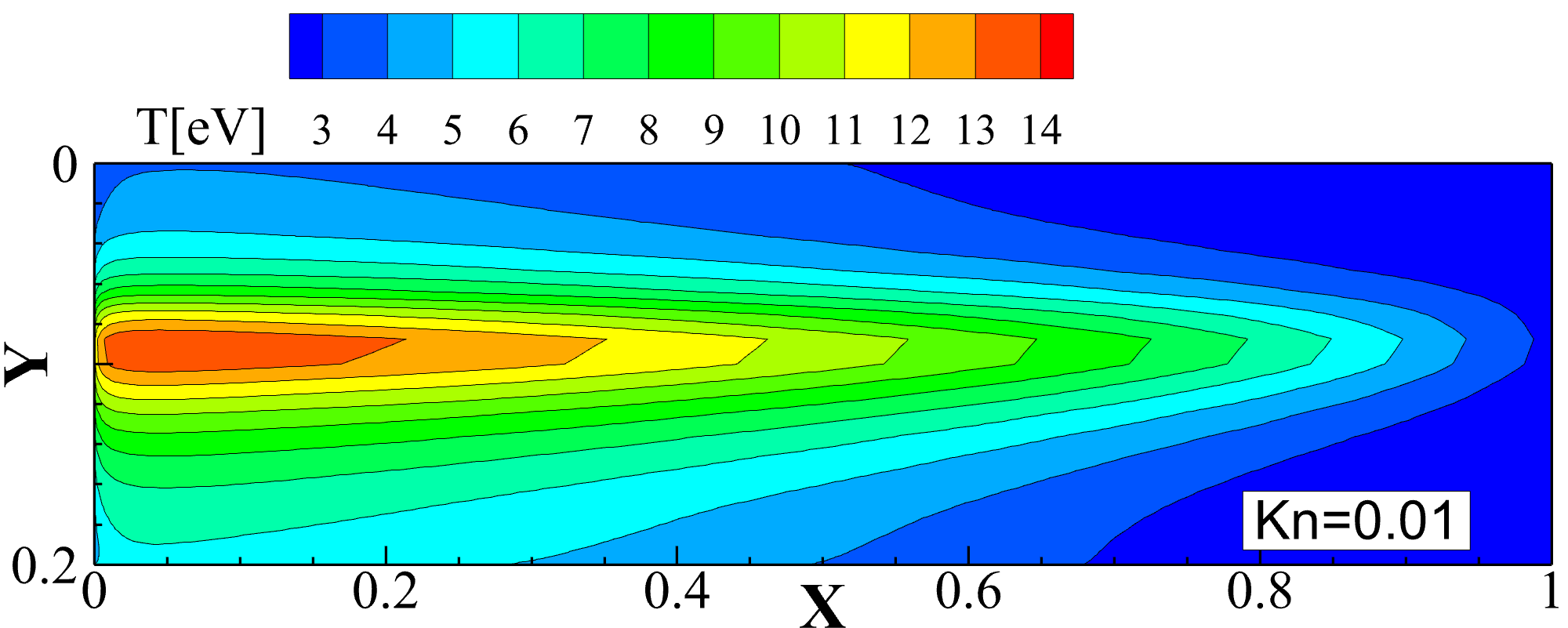}}}
     \subfloat{
{\includegraphics[width=0.45\textwidth,  clip = true]{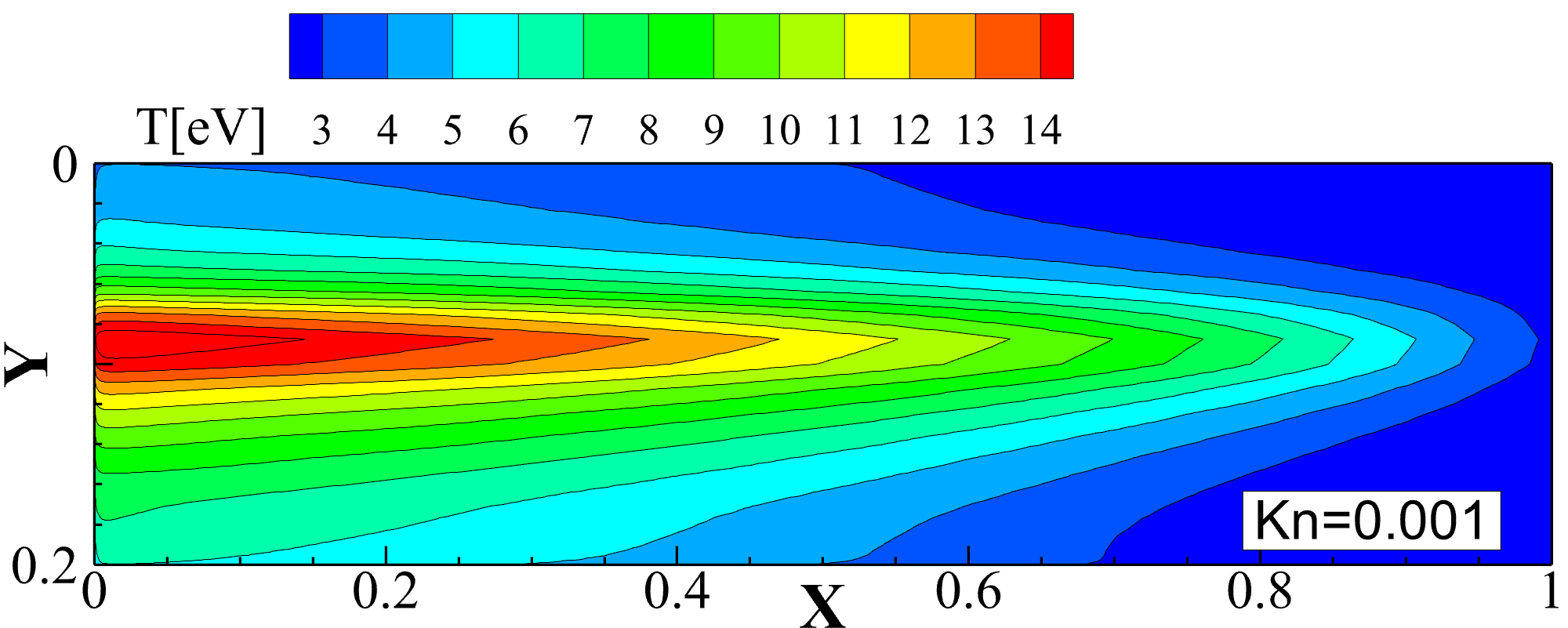}}}
     \caption{Comparisons in the density, temperature, and velocity between the results of GSIS (solid black lines) and CIS (colored background), when Kn=1, 0.1, 0.01, and 0.001. 
     }   
     \label{fig:compare2D}
 \end{figure}

Figure \ref{fig:compare2D} illustrates a comparative analysis of density, velocity, and temperature distributions across a spectrum of Knudsen numbers, spanning from 0.001 to 1. From the density contour plot, it can be observed that when the Knudsen number is small, neutral particles primarily accumulate near the right wall. This occurs for two main reasons. First, when the background ions are reflected from the right wall, some of them are converted into neutral particles, making the right wall effectively a “source” of neutral particles. Second, ions also tend to accumulate in this region, leading to frequent collisions between neutral particles and ions, which significantly slows down the diffusion of neutral particles toward the left outlet. As the Knudsen number increases, although the wall “source” remains relatively constant, the collision frequency between neutral particles and ions decreases sharply, allowing neutral particles to flow quickly from the right to the left outlet, thereby gradually diminishing the accumulation of neutral particles. 
From the velocity contour plot, it can be observed that when the Knudsen number is small, there is a steep velocity gradient near the outlet. Over a very short distance, the velocity rises sharply from nearly zero to close to 25,000 m/s. This is mainly due to the intense collisions with the background particles, which impede the flow of neutral particles. Near the outlet, however, the high degree of vacuum leads to a higher local Knudsen number. The substantial pressure difference over this short distance creates a large pressure gradient, resulting in a very high velocity. As the Knudsen number increases, collisions become less frequent, thus exerting less resistance on the flow of neutral particles. Consequently, the pressure difference acts over a longer distance, resulting in a smaller pressure gradient, and the velocity variation near the outlet becomes relatively gradual.
From the temperature contour plot, it can be observed that as the Knudsen number decreases, the temperature of the neutral particles increasingly aligns with the ion temperature distribution. This is mainly because neutral particles gain energy primarily through collisions with ions. As the Knudsen number decreases, collisions become more frequent, allowing for extensive energy exchange between neutral particles and ions. Ultimately, this results in the temperatures of neutral particles and ions converging.

\begin{figure}[t]
	\centering
	\subfloat{{\includegraphics[width=0.33\textwidth, clip = true]{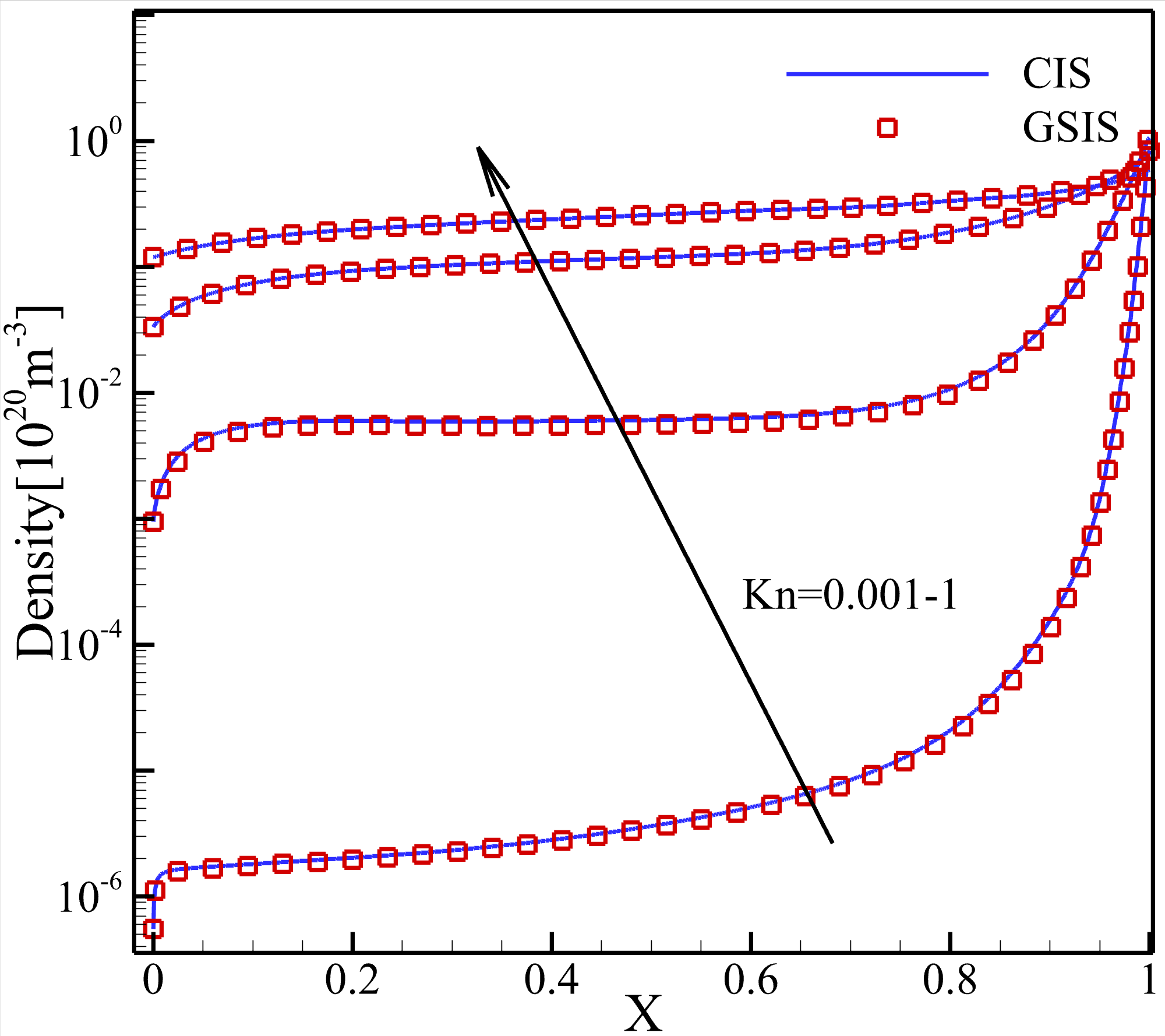}}}
	\subfloat{{\includegraphics[width=0.33\textwidth, clip = true]{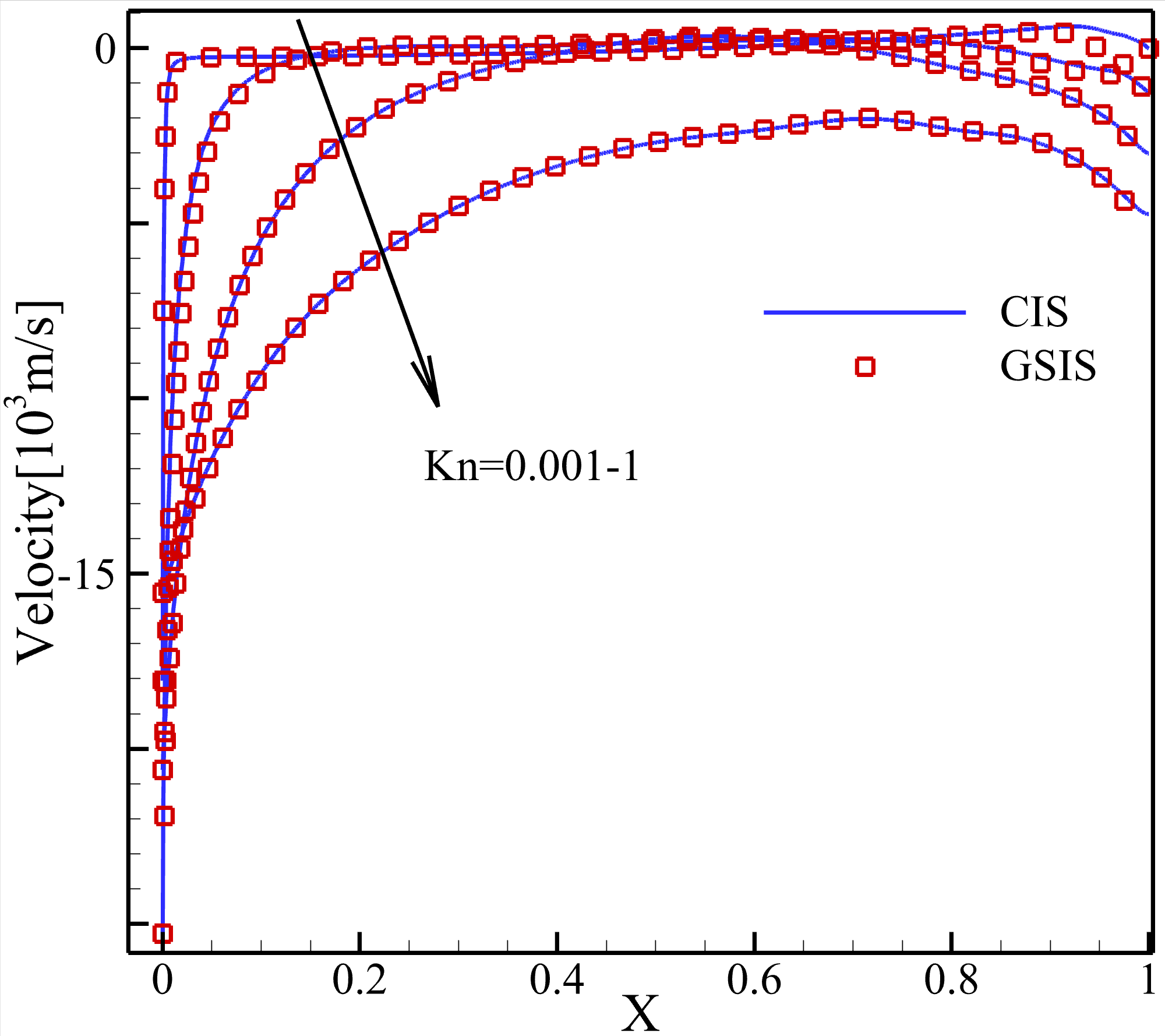}}}
	\subfloat{{\includegraphics[width=0.33\textwidth, clip = true]{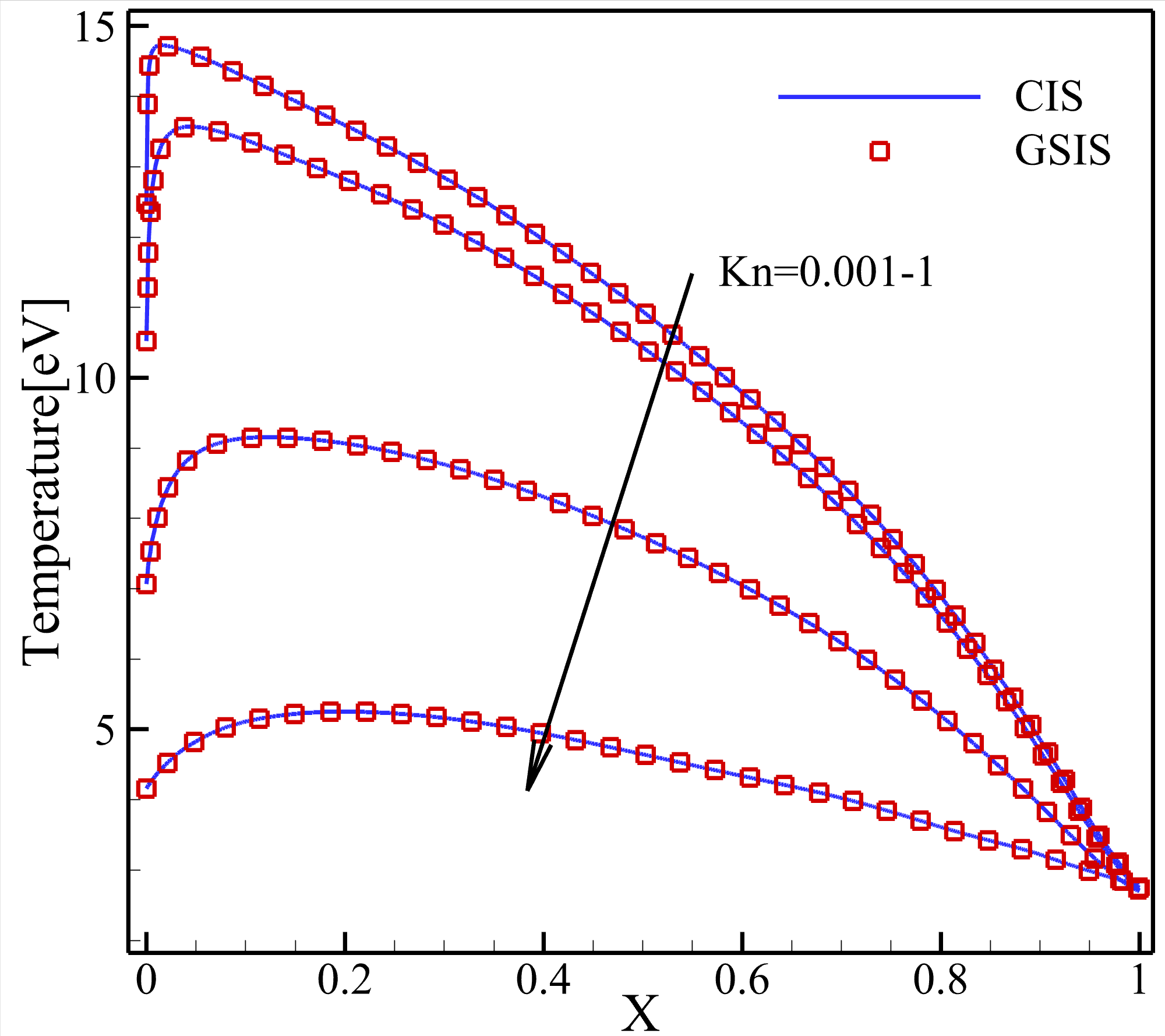}}}
	\caption{
		Comparisons in the density, temperature, and velocity at \(y = 0.1\) between the GSIS (red squares) and CIS (blue solid lines).
	}   
	\label{fig:compare_GSIS2D}
\end{figure}

The striking congruence between the GSIS and CIS outcomes at every Knudsen number underscores the precision and reliability of the GSIS method in depicting the system's behavior under varying levels of rarefaction.
Delving deeper, Fig.~\ref{fig:compare_GSIS2D} offers a detailed examination of the density, velocity, and temperature profiles along the cross-sectional plane at $Y = L_y / 2$. The GSIS method not only matches the benchmark for accuracy but also maintains a consistent correlation with the reference solution across the entire range of Knudsen numbers. This alignment convincingly illustrates the GSIS algorithm's capability to faithfully replicate the reference outcomes delivered by CIS.

Table~\ref{tab:compare_GSIS1} compares the computational costs of CIS and GSIS. Clearly, when $\text{Kn} = 1$, due to the higher rarefaction, the CIS method can quickly reach the steady-state solution, and thus the acceleration effect of GSIS is not significant. This is because, at high rarefaction, the computational demand of the system is already very low, making the traditional CIS method quite efficient. However, as the Knudsen number decreases, the acceleration effect of GSIS increases significantly, demonstrating its unique advantage under low rarefaction conditions. When $\text{Kn} = 0.1$, the acceleration effect in terms of the number of iterations is about 4.4 times, and in terms of time, it is about 2.7 times. When $\text{Kn} = 0.01$, the acceleration effects for both iterations and time increase to approximately 11.9 times and 6.8 times, respectively. When the Knudsen number further decreases to $\text{Kn} = 0.001$, the number of iterations for the CIS method increases sharply, while GSIS achieves an acceleration of 26.8 times in terms of iterations and 18.7 times in terms of time. This shows that under conditions close to the hydrodynamic limit, GSIS can significantly reduce the number of iterations and computational cost through efficient information transfer and coupling acceleration strategies.

\begin{table}[h]
    \centering
    \caption{The convergence steps and computation times of the GSIS and CIS methods at different Knudsen numbers. All simulations are performed on 40 CPU cores.} 
    \begin{tabular}{ccccccc}
        \hline 
        \multirow{2}{*}{\text{Kn}} & \multicolumn{2}{c}{CIS} & \multicolumn{2}{c}{GSIS} & \multicolumn{2}{c}{Acceleration ratio}\\
        \cline{2-7} 
        & Steps & Time (s) & Steps & Time (s) & in Steps  & in Time \\
        \hline 
        0.001 & 2114 & 524 & 79 & 28 &26.8 &18.7\\
        0.01 & 743 & 178 & 62 & 25 &11.9 &6.8\\
        0.1 & 189 & 51 & 43 & 19 &4.4 &2.7\\
        1 & 68 & 20 & 38 &17  &1.7 &1.2\\
        \hline
    \end{tabular}
    \label{tab:compare_GSIS1}
\end{table}

\section{Conclusions}\label{Sec:conclusion}

The general synthetic iterative scheme has been successfully expanded to address plasma edge flows within an implicit finite volume framework. This methodology capitalizes on the synergistic iteration between the macroscopic synthetic equations and the mesoscopic kinetic equation. The computationally demanding kinetic equation supplies high-order constitutive relations for the synthetic equations, which in turn can be swiftly resolved using sophisticated computational fluid dynamics techniques, thereby steering the evolution of the velocity distribution function within the kinetic equations. Consequently, in two demanding numerical test cases, it has been observed that GSIS markedly enhances the computational efficiency of steady-state solutions in comparison to CIS, especially in near-continuum flow regimes. Furthermore, GSIS exhibits asymptotic preserving properties, enabling the use of spatial cell sizes significantly larger than the mean free path of the near-continuum flow, while also displaying a substantial reduction in numerical dissipation compared to CIS. By harnessing its swift convergence and asymptotic preserving attributes, GSIS demonstrates superior efficiency and precision in contrast to CIS. Although our current research is centered on the deterministic solver, the underlying concept can be extended to enhance the DSMC solver for plasma edge flows, as evidenced in an analogous work~\cite{luo_aia}.

\bibliographystyle{elsarticle-num}
\bibliography{ref}

\end{document}